\documentclass[twocolumn,aps,prl,reprint, superscriptaddress,longbibliography]{revtex4-2}

\usepackage{amsmath}
\usepackage{amssymb}
\usepackage{graphicx}
\usepackage{color}
\usepackage{hyperref}
\usepackage[caption=false]{subfig}
\usepackage[colorinlistoftodos]{todonotes}
\usepackage{lipsum}
\usepackage{booktabs}
\usepackage{todonotes}
\usepackage{titlesec}
\usepackage{dcolumn}
\usepackage{bm}

\newcommand{\ket}[1]{\left|#1\right\rangle}
\newcommand{\bra}[1]{\left\langle#1\right|}

\graphicspath{ {main_figures/} {supplement_figures/} }

\makeatletter
\@ifundefined{textcolor}{}
{%
 \definecolor{BLACK}{gray}{0}
 \definecolor{WHITE}{gray}{1}
 \definecolor{RED}{rgb}{1,0,0}
 \definecolor{GREEN}{rgb}{0,1,0}
 \definecolor{BLUE}{rgb}{0,0,1}
 \definecolor{CYAN}{cmyk}{1,0,0,0}
 \definecolor{MAGENTA}{cmyk}{0,1,0,0}
 \definecolor{YELLOW}{cmyk}{0,0,1,0}
}

\makeatother

\begin{document}
\titleformat{\section}{\bfseries\small\centering}{\thesection.}{1em}{\MakeUppercase}

\title{Monitoring fast superconducting qubit dynamics using a neural network}

\author{G.~Koolstra}
\email{gkoolstra@lbl.gov}
\affiliation{Quantum Nanoelectronics Laboratory, Department of Physics, University of California at Berkeley, Berkeley, CA 94720, USA}
\affiliation{Computational Research Division, Lawrence Berkeley National Lab, Berkeley, CA 94720, USA}

\author{N.~Stevenson}
\affiliation{Quantum Nanoelectronics Laboratory, Department of Physics, University of California at Berkeley, Berkeley, CA 94720, USA}

\author{S.~Barzili}
\author{L.~Burns}
\affiliation{Institute for Quantum Studies, Chapman University, Orange, CA 92866, USA}
\affiliation{Schmid College of Science and Technology, Chapman University, Orange, CA 92866, USA}

\author{K.~Siva}
\affiliation{Quantum Nanoelectronics Laboratory, Department of Physics, University of California at Berkeley, Berkeley, CA 94720, USA}

\author{S.~Greenfield}
\affiliation{Institute for Quantum Studies, Chapman University, Orange, CA 92866, USA}
\affiliation{Department of Physics and Astronomy, University of Southern California, Los Angeles CA 90089, USA}

\author{W.~Livingston}
\affiliation{Quantum Nanoelectronics Laboratory, Department of Physics, University of California at Berkeley, Berkeley, CA 94720, USA}

\author{A.~Hashim}
\affiliation{Quantum Nanoelectronics Laboratory, Department of Physics, University of California at Berkeley, Berkeley, CA 94720, USA}
\affiliation{Computational Research Division, Lawrence Berkeley National Lab, Berkeley, CA 94720, USA}

\author{R.~K.~Naik}
\affiliation{Quantum Nanoelectronics Laboratory, Department of Physics, University of California at Berkeley, Berkeley, CA 94720, USA}
\affiliation{Computational Research Division, Lawrence Berkeley National Lab, Berkeley, CA 94720, USA}

\author{J.~M.~Kreikebaum}
\affiliation{Quantum Nanoelectronics Laboratory, Department of Physics, University of California at Berkeley, Berkeley, CA 94720, USA}
\affiliation{Materials Sciences Division, Lawrence Berkeley National Lab, Berkeley, CA 94720, USA}
\author{K.~P.~O'Brien}
\affiliation{Department of Electrical Engineering and Computer Science, Massachusetts Institute of Technology, Cambridge, MA 02139, USA}

\author{D.~I.~Santiago}
\affiliation{Quantum Nanoelectronics Laboratory, Department of Physics, University of California at Berkeley, Berkeley, CA 94720, USA}
\affiliation{Computational Research Division, Lawrence Berkeley National Lab, Berkeley, CA 94720, USA}

\author{J.~Dressel}
\affiliation{Institute for Quantum Studies, Chapman University, Orange, CA 92866, USA}
\affiliation{Schmid College of Science and Technology, Chapman University, Orange, CA 92866, USA}

\author{I.~Siddiqi}
\affiliation{Quantum Nanoelectronics Laboratory, Department of Physics, University of California at Berkeley, Berkeley, CA 94720, USA}
\affiliation{Computational Research Division, Lawrence Berkeley National Lab, Berkeley, CA 94720, USA}
\affiliation{Materials Sciences Division, Lawrence Berkeley National Lab, Berkeley, CA 94720, USA}

\date{\today}

\begin{abstract}
    Weak measurements of a superconducting qubit produce noisy voltage signals that are weakly correlated with the qubit state. To recover individual quantum trajectories from these noisy signals, traditional methods require slow qubit dynamics and substantial prior information in the form of calibration experiments. Monitoring rapid qubit dynamics, e.g. during quantum gates, requires more complicated methods with increased demand for prior information. Here, we experimentally demonstrate an alternative method for accurately tracking rapidly driven superconducting qubit trajectories that uses a Long Short-Term Memory (LSTM) artificial neural network with minimal prior information. Despite few training assumptions, the LSTM produces trajectories that include qubit-readout resonator correlations due to a finite detection bandwidth. In addition to revealing rotated measurement eigenstates and a reduced measurement rate in agreement with theory for a fixed drive, the trained LSTM also correctly reconstructs evolution for an unknown drive with rapid modulation. Our work enables new applications of weak measurements with faster or initially unknown qubit dynamics, such as the diagnosis of coherent errors in quantum gates.
\end{abstract}

\maketitle

\section{Introduction}
Weak measurements allow the observer to reconstruct the dynamics of a quantum system, and to track the evolution of a wavefunction before its collapse to an eigenstate. For superconducting quantum circuits in particular, reconstructing individual quantum trajectories \cite{Gardiner1985,barchielli1986measurement,diosi1988continuous,belavkin1992quantum,CarmichaelPRL1993, WisemanMilburnTrajectories1993, korotkov2001selective,GambettaTrajectories2008,Korotkov2011-qbayes,Korotkov2016} has served as a tool to monitor quantum jumps \cite{Vijay2011-vn,Hatridge2013,Sun2014-zj,Vool2016-de}, track diffusion statistics \cite{Murch2013,Campagne-IbarcqPRX2016,Ficheux2018, Hacohen-Gourgy2016}, generate entanglement via measurement \cite{Roch2014,Chantasri2016-entangledstats}, coherently control quantum evolution using feedback \cite{Vijay2012-rabistable,de2014reversing,Minev2019-gu,Martin2020,Hacohen-Gourgy_and_Martin}, and implement continuous quantum error correction \cite{Mohseninia2020,Atalaya2020,ChenPRR2020, livingston2021}. In weak measurements with superconducting qubits coupled to a readout resonator, the dynamics of the latter are typically much faster than the former. Therefore, the field exiting the resonator is a direct measure of the qubit dynamics, which can be readily computed using quantum filters built on generalizations of Bayes' rule \cite{Korotkov2011-qbayes,Korotkov2016}, Markovian stochastic master equations \cite{JacobsAndSteck2006,GambettaTrajectories2008}, or stochastic path integrals \cite{Chantasri2013-action}. However, such quantum filters are not well-suited to cases where the qubit dynamics are necessarily fast, such as during a rapid entangling gate or syndrome measurement in an error-correction scheme \cite{livingston2021}. Interpreting the output cavity field in this non-Markovian regime for the qubit is challenging. 

A quantum filter obtained by training a Long Short-Term Memory (LSTM) neural network \cite{FlurinPRX2020} is an attractive alternative for reconstructing trajectories in more challenging regimes. An LSTM is a variant of recurrent neural network with persistent memory, which is well suited for time series data with long temporal correlations \cite{Graves2013}. They have recently found applications in quantum physics \cite{CarleoRMP2019} ranging from quantum state tomography \cite{HibatPRR2020, Carrasquilla2019} to qubit noise spectroscopy \cite{WisePRXQ2021, Niu2019Arxiv}. The first application of an LSTM quantum filter to weak measurements successfully tracked slower qubit dynamics with a fidelity comparable to a standard quantum filter \cite{FlurinPRX2020}. We build upon this initial success to show that  LSTM-based filters are not constrained to slow qubit dynamics, but can learn rapid and non-trivial dynamics robustly without additional prior information (e.g. from calibration measurements) which would be required by traditional methods. 

\begin{figure}
    \centering
    \includegraphics[width=\columnwidth]{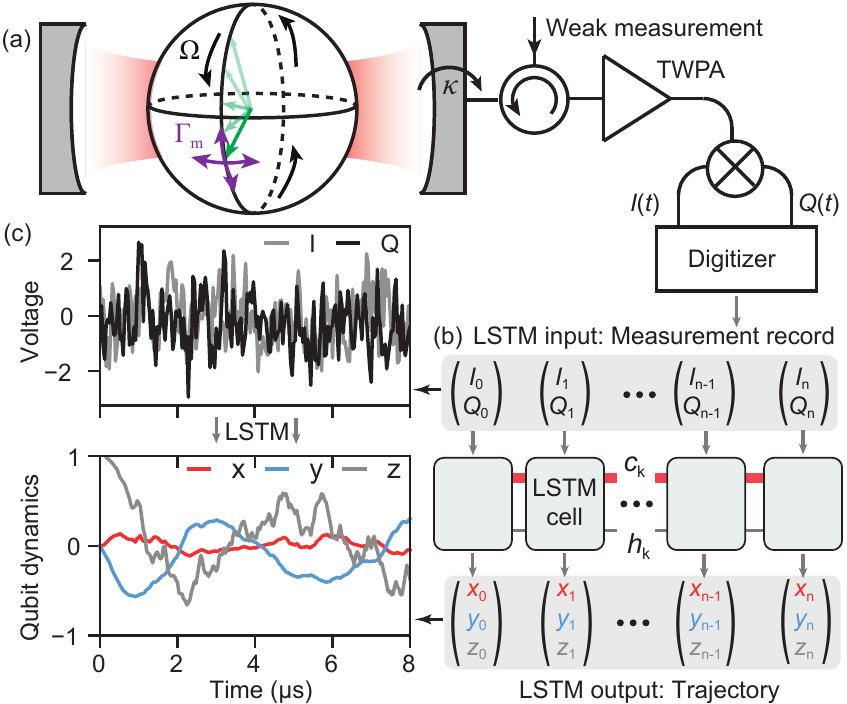}
    \caption{Monitoring fast qubit dynamics using an LSTM artificial neural network. (a) During weak measurement of a strongly driven qubit, the qubit state (green arrow) evolves due to a coherent Rabi drive (black, rate $\Omega$) and backaction from both quadratures of a heterodyne measurement (purple arrows). If $\Omega$ exceeds the resonator relaxation rate $\kappa/2$, memory effects such as delay and qubit-resonator correlations can no longer be ignored when reconstructing trajectories. (b) The LSTM learns the stochastic, dissipative qubit dynamics, mapping elements of each noisy voltage record ($I, Q$) to a qubit Bloch vector trajectory ($x, y, z$). The cell state $c_k$ and hidden state $h_k$ can encode long-term correlations in the measurement record. (c) Sample record with corresponding qubit trajectory output by the trained LSTM.}
    \label{fig:fig1}
\end{figure}

\begin{figure*}
    \centering
    \includegraphics[width=0.75\textwidth]{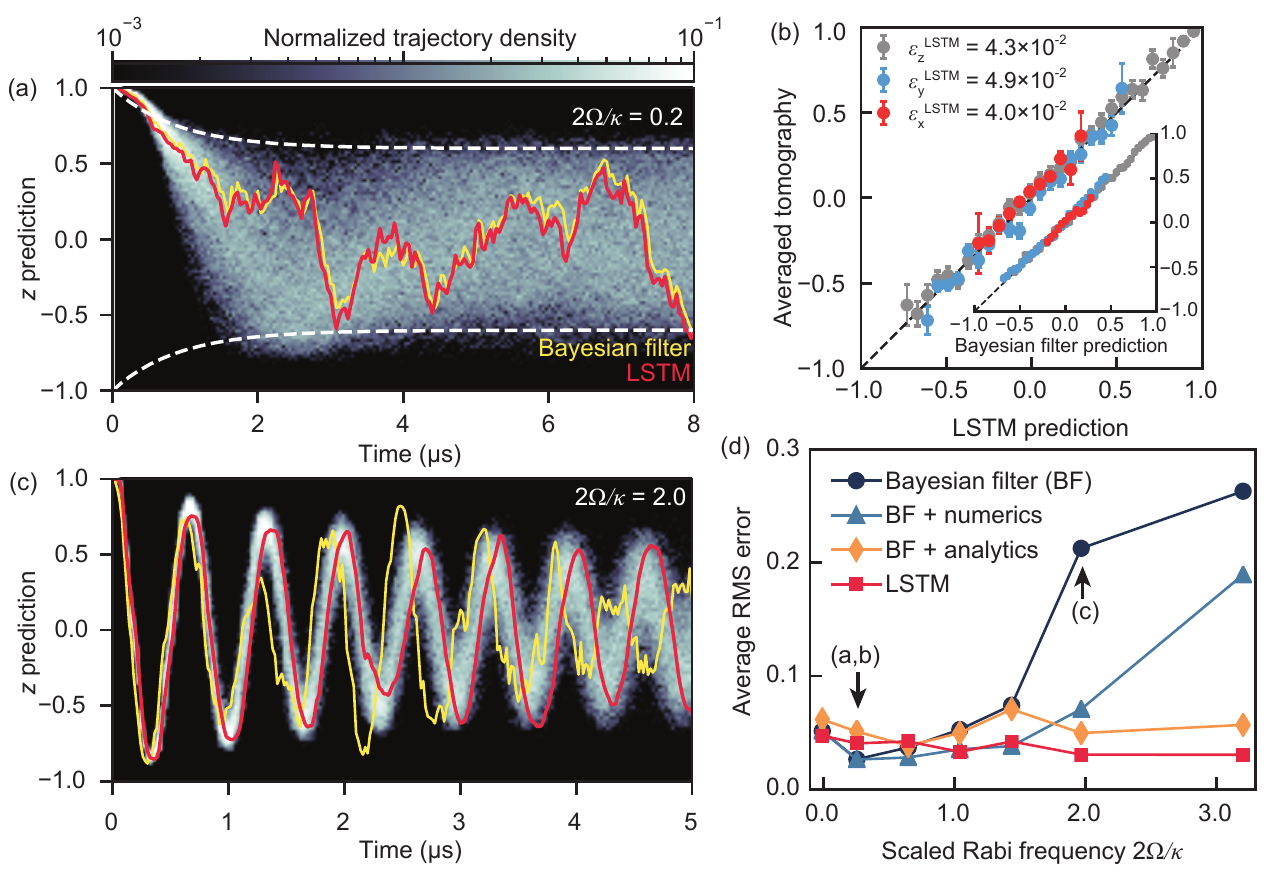}
    \caption{Breakdown of adiabatic trajectory reconstruction. (a) The histogram of weakly driven trajectories of length $T_m = 8.0$ \textmu{}s (reconstructed by the LSTM with $\mathrm{d}t=40$~ns) shows rapid trajectory diffusion due to measurement backaction. The colorbar represents the trajectory probability density at each timestep. The LSTM produces trajectories (example in red) comparable to those from a steady-state Bayesian filter (yellow). (b) LSTM validation based on tomographic measurements immediately following the LSTM prediction, for the trajectories shown in (a) where $2\Omega/\kappa = 0.2$. The dashed line with slope 1 indicates perfect validation. The inset shows the Bayesian filter validation for the same data set with slightly smaller RMS error $(\varepsilon_x^\mathrm{BF}, \varepsilon_y^\mathrm{BF}, \varepsilon_z^\mathrm{BF}) = (2.3, 3.2, 2.5)\times 10^{-2}$. This is partly because $\varepsilon_{x,y,z}^\mathrm{BF}$ are computed using the entire dataset, whereas $\varepsilon_{x,y,z}^\mathrm{LSTM}$ only have access to trajectories not used for training. (c) For fast qubit dynamics ($2\Omega/\kappa = 2.0$) outside the adiabatic regime the predictions of the steady-state Bayesian filter and LSTM diverge (trajectory $\mathrm{d}t=20$~ns). (d) Validation errors averaged over the three qubit coordinates vs. Rabi frequency $2\Omega/\kappa$, showing a breakdown of the steady-state Bayesian filter for $2\Omega/\kappa > 1$ (dots, BF), while the LSTM validation error (squares) remains small. The performance of the Bayesian filter improves with additional numerical prior information of the expected evolution of the $z$-conditioned resonator fields (BF + numerics, and Appendix~\ref{chap:bsu_corrections}), and improves even further when adding analytical corrections to the measurement backaction (BF + analytics, and Appendix~\ref{chap:tilting_derivation}). Importantly, the LSTM performance stays consistent without additional prior information. Arrows mark the data shown in (a), (b) and (c).}
    \label{fig:fig2}
\end{figure*}

In this work, we demonstrate that an LSTM accurately reconstructs quantum state trajectories of a driven superconducting qubit coupled to a readout resonator, even for qubit dynamics faster than the relaxation timescale of the resonator. Our LSTM trains entirely on experimental observations, compensates for the delays and correlations originating from the limited detection bandwidth, and outperforms conventional reconstruction methods in the case of fast qubit dynamics. By extracting and separating the coherent dynamics and measurement backaction learned by the LSTM, we observe nontrivial drive-dependent corrections consistent with theory that includes the joint dynamics of the qubit and resonator (information not given to the LSTM). The dominant corrections include a rotation of the measured eigenstates and a reduction in the effective measurement rate with increasing drive. Finally, we demonstrate that the trained LSTM can correctly identify and track a priori unknown time-dependencies in the dynamics produced by modulating the drive, despite no prior examples of such behavior in the training data. Our results suggest that LSTM-based filters may enable novel applications of continuous monitoring to previously unobtainable regimes. 

\section{Experimental system}
Our qubit-resonator system (Fig \ref{fig:fig1}a and Appendix~\ref{chap:experimental_setup} and \ref{chap:device_parameters_and_stability}) consists of a superconducting transmon qubit (frequency $\omega_{ge}/2\pi = 5.473$ GHz) capacitively coupled to a superconducting coplanar waveguide resonator ($\omega_\mathrm{res}/2\pi = 6.679$ GHz). The interaction Hamiltonian in this regime is dispersive
\begin{equation}
\label{eqn:interaction-hamiltonian}
H_\text{int} = \hbar \chi a^\dagger a \sigma_z,
\end{equation}
where $\hbar$ is the reduced Planck's constant, $\chi/2\pi = 0.47$ MHz is the qubit-dependent resonator frequency shift, $a^\dagger$ ($a$) is the creation (annihilation) operator for the readout resonator, and $\sigma_{z}$ is the qubit Pauli $z$-operator. We additionally drive the qubit (at its Stark-shifted transition frequency) to induce Rabi oscillations about the $x$-axis at a variable Rabi frequency $\Omega$.

To continuously monitor this system we apply a weak probe tone to the resonator at the frequency midpoint between qubit-shifted resonances, which populates the resonator with an approximate mean photon number of $\bar{n} \approx 0.3$. After interacting with the qubit, the resonator field escapes the resonator at a rate $\kappa/2 \approx 2\pi \times 0.8$ MHz. We amplify both in-phase ($I$) and quadrature ($Q$) parts of this field using a traveling wave parametric amplifier (TWPA), perform a heterodyne measurement of its complex amplitude, and digitize the resulting pair of noisy quadrature voltages with a 1 ns sampling time. After a chosen monitoring duration $0 < T_m < 8$ \textmu{}s of variable length (Fig.~\ref{fig:fig1}c) we turn off the Rabi drive and projectively measure the qubit along one of the cardinal axes of the Bloch sphere ($\sigma_{x,y,z}$). We use this final projective measurement as a training label and for verification via conditioned final state tomography.

Our phase-preserving detection technique \cite{Campagne-IbarcqPRX2016, Steinmetz2021} gives rise to two distinct types of measurement backaction: nonunitary partial collapse from information that distinguishes the qubit states, and unitary phase drift from fluctuations in resonator photon number \cite{Korotkov2016}. Our detection technique is robust to drifts in amplifier gain, spans several GHz, and in contrast to phase-sensitive detection of just the informational quadrature \cite{FlurinPRX2020}, allows for simultaneous weak measurement of multiple qubits \cite{Roch2014} in future experiments.

Before passing a monitoring record into a quantum filter, we digitally filter it to remove qubit-independent high-frequency noise, then coarse-grain it into longer time bins while ensuring rapidly driven qubit dynamics is not undersampled. We then pass the coarse-grained digital record $\{(I_k,Q_k)\}_{k=0}^n$ into both the LSTM filter and a more traditional Bayesian quantum filter, and compare the resulting qubit trajectories expressed in the Bloch coordinates $\{(x_k,y_k,z_k)\}_{k=0}^n$.

We train the LSTM (Fig.~\ref{fig:fig1}b) to reconstruct trajectories by feeding it a subset of experimental records $\{(I_k, Q_k)\}_{k=0}^n$ reserved for training, which are of varying length $T_m$ and labeled by their final projective measurement basis and outcome. During training the LSTM adjusts its network parameters to minimize the cross-entropy loss, which is computed from predictions at $t = T_m$ and associated projective measurement outcomes (for details see Appendix~\ref{chap:network_training}). We have found that the LSTM network converges better than a simple feedforward neural network, which we attribute to the LSTM's ability to track long correlations in the cell state. After the LSTM converges we feed the trained model individual measurement records (Fig.~\ref{fig:fig1}c) from separate data sets not used for training. In the rest of this work, we analyze the trajectories $\{(x_k, y_k, z_k)\}_{k=0}^n$ predicted by the trained network from such test data.

\section{Qubit trajectory reconstruction with a neural network}
Qubit dynamics faster than the resonator relaxation rate $\kappa/2$ can prevent the conditioned coherent steady-states in the resonator from adiabatically following their associated qubit states \cite{Korotkov2016}. To investigate the onset of this non-adiabatic behavior beyond the coherent-state approximation \cite{GambettaTrajectories2008,Korotkov2016}, we experimentally vary the rate of the qubit dynamics by tuning Rabi frequency $\Omega$ from slow dynamics ($2\Omega/\kappa \ll 1$) to fast dynamics that are well outside the adiabatic regime ($2\Omega/\kappa > 1$), training the LSTM to predict trajectories for each Rabi frequency independently. 

We first use the LSTM to reconstruct qubit dynamics with a weak Rabi drive ($2\Omega/\kappa = 0.2$), where conventional steady-state methods can still accurately reconstruct trajectories. A histogram of reconstructed trajectories (Fig.~\ref{fig:fig2}a) shows oscillations due to the Rabi drive as well as diffusion due to measurement backaction, which slowly collapses the trajectories towards $\ket{\pm Z}$. The competition between (i) the Rabi drive, (ii) trajectory collapse and diffusion at rate $\eta \Gamma_m$, and (iii) trajectory dephasing at rate $2(1-\eta)\Gamma_m$ confines trajectories to a Bloch sphere with reduced radius (dashed lines in Fig.~\ref{fig:fig2}a, and Appendix~\ref{chap:traj_histograms}) set by the measurement efficiency $\eta$. Here, $\Gamma_m$ is defined as the single quadrature measurement dephasing rate which is set by the amplitude of the weak measurement tone.

When comparing individual trajectories, the LSTM produces trajectories similar to a conventional Bayesian filter approach. This Bayesian filter method sequentially estimates the quantum state from a measurement record using a well-known initial state and calibrated values for $\Omega$, the ensemble decay rate and qubit-conditioned resonator output fields. We quantify the error of both methods self-consistently by averaging projective measurement results of trajectories with similar predictions \cite{Ficheux2018,Campagne-IbarcqPRX2016,FlurinPRX2020}. The averaged tomography results closely follow the LSTM predictions for all three Bloch coordinates (Fig.~\ref{fig:fig2}b) with RMS error $(\varepsilon_x^\mathrm{LSTM}, \varepsilon_y^\mathrm{LSTM}, \varepsilon_z^\mathrm{LSTM}) = (4.0, 4.9, 4.3) \times 10^{-2}$ (definition in Eq.~\eqref{eq:supp_epsilon_definition}). This error largely reflects our imperfect knowledge of the true quantum state from a finite number of projective measurements, see Appendix~\ref{chap:rnn_accuracy}. More training data is likely to improve both the estimate of the true quantum state and the LSTM accuracy, since the LSTM training loss does not saturate when training on increasing fractions of available training data. Nevertheless, even with a finite size training dataset considered here ($\sim4\times10^5$ voltage records), the LSTM learns an accurate representation of the reduced qubit dynamics, with a lower bound on the accuracy of $1 - \varepsilon_\mathrm{avg}^\mathrm{LSTM}/2 = 0.978$.

For large Rabi frequencies the LSTM trajectories remain faithful, even when the qubit dynamics exceeds the relaxation rate of the resonator $\kappa/2$ (Figs.~\ref{fig:fig2}c,d). In contrast, the Bayesian filter's validation error increases sharply past  $2\Omega/\kappa \approx 1$, because this method is sensitive to two errors related to the sampling time of voltage records, $\mathrm{d}t$. The first error, due to stepwise application of fast coherent dynamics, grows with $\Omega \,\mathrm{d}t$ \cite{Rouchon2015}, whereas the second error due to temporal correlations grows for $\mathrm{dt} < 2 / \kappa$ \cite{WisemanMilburn2009}. Consequently, for fast Rabi frequencies $2\Omega/\kappa > 1$ there is no choice of $\mathrm{d}t$ for which the standard Bayesian filter produces accurate trajectories.  

Experimentally the breakdown of the Bayesian filter coincides with a notable delay between oscillations in the measured voltage records and the qubit coordinate $z(t)$ (see Fig.~\ref{fig:supp_comp_volt_rec_trajs}), which occurs because the resonator retains photons for a time $\kappa^{-1}$. In addition, the coherent oscillation amplitude in the voltage record reduces significantly for $2\Omega/\kappa > 1$, because instead of measuring the instantaneous qubit state, photons measure the time-averaged measurement operator \cite{SzombatiPRL2020}. This effect is similar to the suppression of current oscillations in a semiconducting point contact detector \cite{StaceBarrett2004}.

As a first order correction of the resonator memory effects, we adjust the Bayesian filter based on independent master equation simulations of the resonator field amplitudes conditioned on the qubit states (Fig.~\ref{fig:fig2}d, BF + numerics and further details in Appendix~\ref{chap:bsu_corrections}). While this reduces the RMS error at moderate Rabi frequencies, the average error remains large for $2\Omega/\kappa > 1$. Only after an analytic treatment of the resonator memory, and including calibrated parameters such as $\chi$ and  $\kappa$, does the Bayesian filter reach similar error as the LSTM for all Rabi frequencies (Fig.~\ref{fig:fig2}d, BF + analytics and further details in Appendix~\ref{chap:tilting_derivation})). This indicates that with enough prior information the Bayesian filter with analytic correction reproduces the essential ingredients of the LSTM trajectories. However, the LSTM offers an accurate reconstruction method requiring no prior knowledge of coupling rates to the environment.

\section{Resonator memory corrections to qubit trajectories}
\begin{figure*}
    \centering
    \includegraphics[width=\textwidth]{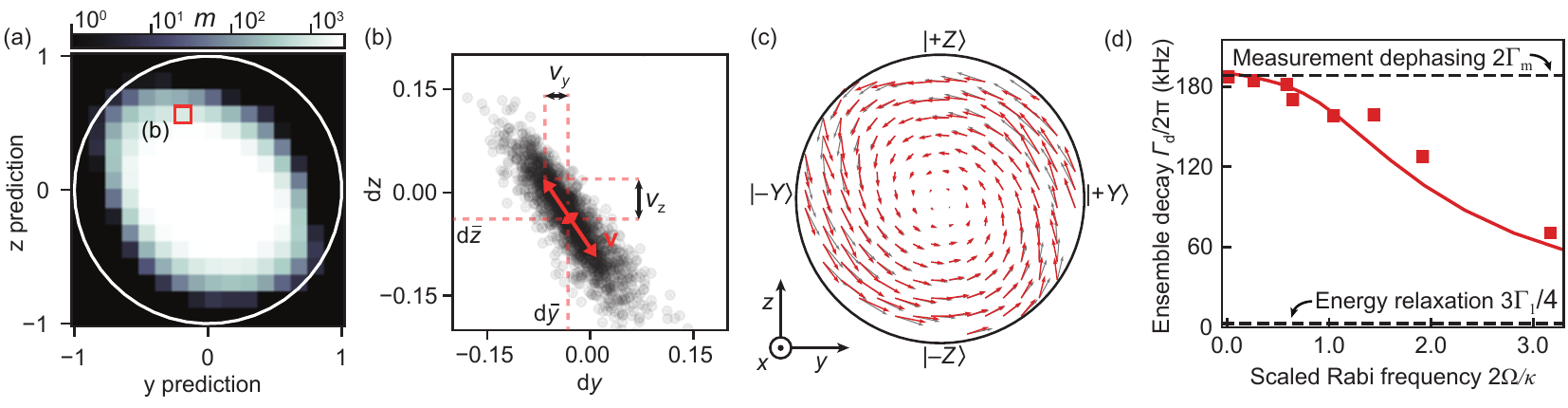}
    \caption{LSTM trajectory processing and unraveling Lindbladian trajectory dynamics. (a) Histogram of the LSTM trajectories in the $yz$-plane of the Bloch sphere for $2\Omega / \kappa = 0.6$. To extract trajectory drift and diffusion, we perform statistics on the Bloch vector increments $\{(\mathrm{d}y_k, \mathrm{d}z_k)\}_{k=0}^m$, where $m$ is the number of samples in each bin, see colorbar. (b) Statistics of the Bloch vector increments for the highlighted pixel in (a). The mean of the Bloch vector increments gives the average drift  ($\mathrm{d}\bar{y}, \mathrm{d}\bar{z}$), whereas the scaled eigenvector of the covariance matrix ($\mathbf{v}$) gives the direction and magnitude of stochastic kicks due to measurement backaction. We compute ($\mathrm{d}\bar{y}, \mathrm{d}\bar{z}$) and $\mathbf{v}$  for each pixel in the $yz$-plane to visualize the Lindbladian dynamics and measurement backaction. (c) The average drift $(\mathrm{d}\bar{y}, \mathrm{d}\bar{z})$ of trajectories binned in the $yz$-plane ($2\Omega/\kappa = 0.6$, red arrows) reveals dynamics consistent with the applied Hamiltonian $H = \Omega \sigma_x/2$ and decay towards $y = 0$ at a rate $\Gamma_d$. The gray arrows represent a fit of the extracted pairs $(\mathrm{d}\bar{y}, \mathrm{d}\bar{z})$ to Eqs.~\eqref{eq:avg_drift_dy}-\eqref{eq:avg_drift_dz}, with $\Omega$ and $\Gamma_d$ as free parameters. (d) The decay rate $\Gamma_d/2\pi$ (red squares) falls from the expected measurement dephasing rate $2\Gamma_m$ towards the bare qubit relaxation rate \cite{Yan2013}, as the qubit and resonator decouple. A master-equation simulation of the qubit-resonator system (solid line) agrees qualitatively with the LSTM trajectories.}
    \label{fig:fig3}
\end{figure*}

To gain further insight into why the LSTM outperforms the Bayesian filter for $2\Omega/\kappa > 1$, we perform statistical analysis on the LSTM trajectories. First, we bin the LSTM trajectories inside the Bloch sphere (Fig.~\ref{fig:fig3}a) and calculate the Bloch vector increments $\mathrm{d}\mathbf{r} = \mathbf{r}_{t+1} - \mathbf{r}_{t}$ for each bin, where $\mathbf{r} = (x,y,z)$. Since the Rabi drive confines most of the interesting qubit dynamics to the $yz$-plane, we restrict our analysis to two dimensions. Therefore, each bin contains a set of Bloch vector increments $\{(\mathrm{d}y_k, \mathrm{d}z_k)\}_{k=0}^m$, where each pair can be seen as a random variable due to the stochastic nature of the measurement backaction (Fig.~\ref{fig:fig3}b, dots). Next, we compute the average drift $(\mathrm{d}\bar{y}, \mathrm{d}\bar{z})$ and diffusion from the sample mean and covariance matrix on $\{(\mathrm{d}y_k, \mathrm{d}z_k)\}_{k=0}^m$, respectively. The coviarance matrix
\begin{align}
	C(y, z) &= \begin{pmatrix} \sigma_{\mathrm{d}y}^2 & \sigma_{\mathrm{d}y,\mathrm{d}z} \\  \sigma_{\mathrm{d}y,\mathrm{d}z} & \sigma_{\mathrm{d}z}^2 \end{pmatrix} \label{eq:cov_matrix} \\
	 \sigma_{\mathrm{d}y}^2 &= \frac{1}{m} \sum_{k=0}^m (\mathrm{d}y_k - \mathrm{d}\bar{y})^2 \\
	 \sigma_{\mathrm{d}y,\mathrm{d}z} &= \frac{1}{m} \sum_{k=0}^m (\mathrm{d}y_k - \mathrm{d}\bar{y})(\mathrm{d}z_k - \mathrm{d}\bar{z}) \\ 
	 \sigma_{\mathrm{d}z}^2 &= \frac{1}{m} \sum_{k=0}^m (\mathrm{d}z_k - \mathrm{d}\bar{z})^2
\end{align}
is convenient to quantify diffusion since its largest eigenvalue $\lambda_\mathrm{max}$ represents the magnitude of the backaction, and the associated eigenvector $\mathbf{\xi}$ indicates the direction of the backaction. The vector $\mathbf{v} = (v_y, v_z) = \lambda_\mathrm{max} \mathbf{\xi}$ thus quantifies diffusion magnitude and direction, and is visualized in Fig.~\ref{fig:fig3}b. By separating the drift and diffusion as described above, we can separately compare the average drift to Lindbladian dynamics, and the diffusion to the measurement backaction dynamics. 

In the limit of a weak Rabi drive, the expected drift and diffusion can be directly calculated from the Stochastic master equation in Eqs.\eqref{eq:sme_het_dx}-\eqref{eq:sme_het_dz}. Therefore, in the plane of the Rabi drive ($x = 0$), we expect the average drift to take on the form
\begin{align}
	\mathrm{d} \bar{y} &= -\Gamma_d y \mathrm{d} t + \Omega z \mathrm{d}t \label{eq:avg_drift_dy} \\ 
	\mathrm{d} \bar{z} &= - \Omega y \mathrm{d} t \label{eq:avg_drift_dz}.
\end{align}
Similarly, the expected diffusion is
\begin{align}
	\mathrm{Var} (\mathrm{d} y) &= -2 \eta \Gamma_m zy \mathrm{d} t \label{eq:diffusion_dy} \\
	\mathrm{Var} (\mathrm{d} z) &= 2 \eta \Gamma_m (1 - z^2) \mathrm{d} t.\label{eq:diffusion_dz}
\end{align}
Here, $\Gamma_d$ is the ensemble average dephasing, $\Gamma_m$ is the single quadrature measurement rate and $\eta$ is the total measurement efficiency. In the remainder of this section we will first study the effect of the increasing Rabi drive on the average drift, and then study deviations from the backaction model of Eqs.\eqref{eq:diffusion_dy} and \eqref{eq:diffusion_dz}.

In Fig.~\ref{fig:fig3}c we show an example of the average drift for $2\Omega/\kappa = 0.6$. The trajectory drift (red arrows) shows good agreement with the expected drift from the Hamiltonian $H = \Omega/2~\sigma_x$ and dissipation pulling trajectories towards the origin at a rate $\Gamma_d/2\pi$ (gray arrows). From a fit of the average drift to Eqs.~\eqref{eq:avg_drift_dy}-\eqref{eq:avg_drift_dz}, using $\Omega$ and $\Gamma_d$ as free parameters, we find $\Gamma_d/2\pi = 0.188 \pm 0.007$~MHz. For increasing Rabi drive strengths, Eqs.~\eqref{eq:avg_drift_dy}-\eqref{eq:avg_drift_dz} continue to fit the average drifts. Interestingly, we observe a monotonic decrease in $\Gamma_d/2\pi$ (Fig.~\ref{fig:fig3}d), which we attribute to decoupling of the qubit from the weak measurement when the Rabi splitting of the dressed qubit exceeds the cavity linewidth \cite{SzombatiPRL2020}. This decoupling behavior is well captured by a master equation simulation of the qubit-resonator system (Fig.~\ref{fig:fig3}d, solid line). This master equation simulation also correctly shows a small Rabi drive assymptote $\Gamma_d \rightarrow 2\Gamma_m$ due to measurement dephasing \cite{Gambetta2006} and another assymptote $\Gamma_d \rightarrow 3\Gamma_1/4$ \cite{Yan2013} due to natural relaxation, when qubit and resonator are completely decoupled.

\begin{figure}
    \centering
    \includegraphics[width=\columnwidth]{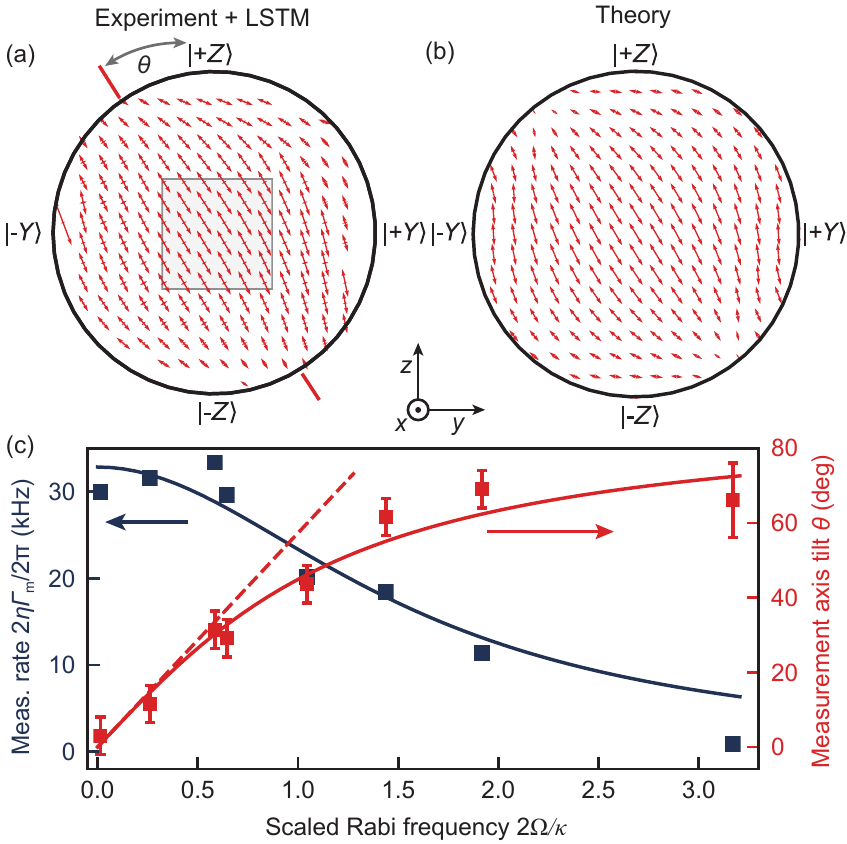}
    \caption{Resonator memory corrections to the measurement backaction from LSTM trajectories. (a) Trajectory diffusion map for $2\Omega/\kappa = 0.6$, obtained by computing the eigenvectors of the covariance matrix Eq.\eqref{eq:cov_matrix} for trajectories binned in the $yz$-plane. See also Figs.~\ref{fig:fig3}a,b. The tilt towards $\ket{\pm Y}$ in the plane of the Rabi drive is in contrast to the prediction from Eq.~\eqref{eq:diffusion_dy}-\eqref{eq:diffusion_dz}. (b) The theory prediction, detailed in Appendix~\ref{chap:supp_deriv_back_map}, includes resonator memory effects and reproduces the tilt. (c) For small $\Omega/\kappa$ the measurement axis tilts linearly with the Rabi frequency $\theta = 2 \Omega / \kappa$ (dashed line). The experimental tilt of the measurement eigenstates in the $yz$-plane (red squares) is accompanied by a decrease in measurement rate (blue squares). Solid lines are fits to Eq.~\eqref{eq:theory_tilt} (red) and Eq. \eqref{eq:reduced_measurement_rate} (blue), respectively. The error bars for $\theta$ are estimated from the imprecision in determining the tilt angle. The error bars for the measurement rate obtained from fitting are smaller than the markers.}
    \label{fig:fig4}
\end{figure}

Since the average drift follows the expected Lindbladian dynamics even for fast Rabi rates, we now shift our attention to the measurement backaction, which we visualize by plotting the scaled eigenvectors $\mathbf{v}$ (Fig.~\ref{fig:fig3}b) of $C(y,z)$. At low Rabi frequencies ($2\Omega / \kappa = 0$), we find both informational backaction, vanishing at the poles $\ket{\pm Z}$ of the Bloch sphere, and phase backaction, consistent with a dispersive heterodyne measurement (see Appendix~\ref{chap:hetdyne_backaction}). Furthermore, a fit of the extracted diffusion $(v_y, v_z)$ to Eqs.~\eqref{eq:diffusion_dy}-\eqref{eq:diffusion_dz}, along with $\Gamma_d/2\pi = 0.175$~MHz determined from Fig.~\ref{fig:supp_efficiency_calibration}d, gives the efficiency of the measurement chain $\eta = 2\Gamma_m / \Gamma_d = 0.185 \pm 0.002$. This value is in agreement with an independent calibration measurement ($\eta = 0.188\pm0.003$ (see Fig.~\ref{fig:supp_efficiency_calibration}). 

For large Rabi drives, we observe two significant corrections to the expected measurement backaction of Eqs.~\eqref{eq:diffusion_dy}-\eqref{eq:diffusion_dz}: a tilt of the measurement poles in the plane of the Rabi drive towards $\ket{\pm Y}$ and a reduced trajectory diffusion rate $2\eta\Gamma_m$ (Fig.~\ref{fig:fig4}a). These corrections complicate extracting the diffusion rate. Therefore, we first determine the rotation $\theta$ from the average angle between $\mathbf{v}$ and the $z$-axis from bins near the origin (Fig.~\ref{fig:fig4}a, shaded square). Next, we apply a coordinate rotation $(y',z') \mapsto (y \cos \theta - z \sin \theta, y\sin \theta + z \cos \theta)$ to undo the observed rotation, which now allows us to fit the diffusion $(v_{y'}, v_{z'})$ to Eqs.~\eqref{eq:diffusion_dy}-\eqref{eq:diffusion_dz}. The fit results for $\theta$ and the measurement rate $2 \eta \Gamma_m$ are shown in Fig.~\ref{fig:fig4}c, squares. Intuitively, the tilt occurs because the Rabi drive drags the qubit state counterclockwise in the $yz$-plane while photons in the resonator measure the qubit $z$ coordinate for a characteristic time $\tau_c$. This simple picture results in a delay between qubit state and measurement record which is proportional to $\Omega\, \tau_c$, consistent with the linear increase in $\theta$ for small Rabi frequencies in Fig.~\ref{fig:fig4}c. 

To explain the effects of the slow measurement beyond small $\Omega/\kappa$, we derive an analytical model for the resonator mode $\hat{a}$ in terms of the Bloch coordinate $z$, taking into account the resonator memory (see Appendix~\ref{chap:tilting_derivation}). For a constant Rabi drive in the $yz$-plane the weak measurement no longer acts purely along $\hat{z}$, but along 
\begin{equation}
    \hat{z}_\mathrm{eff} = \int_0^t z(t-\tau) \mathrm{d}P(\tau) = \sqrt{\eta_\mathrm{avg}} \left[ \cos \theta \, \hat{z} - \sin \theta \, \hat{y} \right], \label{eq:tilted_meas_axes_result}
\end{equation}
where $P$ is the inverse Fourier transform of the Lorentzian resonator spectrum, 
\begin{equation}
	\theta = \arctan(\Omega \tau_c) \label{eq:theory_tilt}
\end{equation} 
is the measurement axis tilt, $\tau_c = \int \tau P(\tau) \mathrm{d}\tau =  2/\kappa$ is the predicted delay time, and $\eta_\mathrm{avg}$ captures the reduction in measurement rate in the $yz$-plane:
\begin{equation}
    \Gamma_m (\Omega) = \eta_\mathrm{avg} \Gamma_m (0) = \frac{\Gamma_m (0)}{1 + \left(\Omega\tau_c \right)^2}. \label{eq:reduced_measurement_rate}
\end{equation}
From a simultaneous fit of both quantities in Fig.~\ref{fig:fig4}c to Eqs.~\eqref{eq:theory_tilt} and \eqref{eq:reduced_measurement_rate}, we find the resonator memory time $\tau_c = 0.20 \pm 0.01$~\textmu{}s, which agrees with the relaxation time $2/\kappa = 0.2$~\textmu{}s from spectroscopic measurements. For Rabi frequencies exceeding $2\Omega/\kappa \approx 2.0$ the rate of trajectory diffusion in the $yz$-plane is greatly suppressed, showing that a fast Rabi frequency protects the qubit from measurement backaction \cite{SzombatiPRL2020}. 

The results of Fig.~\ref{fig:fig3} and \ref{fig:fig4} demonstrate accurate estimation of various decay rates, measurement efficiency and the memory time of the system from a simple trajectory decomposition that requires no prior knowledge of the resonator memory. This makes the LSTM a useful tool in the context of parameter estimation from weak measurements \cite{NaghilooPRL2017, Genois2021}. In addition, the corrections to the measurement backaction shed further light on the breakdown of Baysian filters for trajectory reconstruction. 

\section{Uncovering hidden time-dependencies}
A major advantage of the LSTM is that trajectory reconstruction does not require prior knowledge of system and environment parameters, which may fluctuate or may be hard to calibrate. To highlight this advantage, we perform a new set of weak measurements where we vary the Rabi frequency sinusoidally with a period of 1.8~\textmu{}s, thereby obscuring the prior information necessary for traditional trajectory reconstruction methods. The ability to reconstruct trajectories with a priori unknown time-dependencies is important for a wide range of experiments, such as continuous quantum error correction or weak measurements of a highly non-Markovian environment \cite{Shabani2014}.

\begin{figure}
    \centering
    \includegraphics[width=\columnwidth]{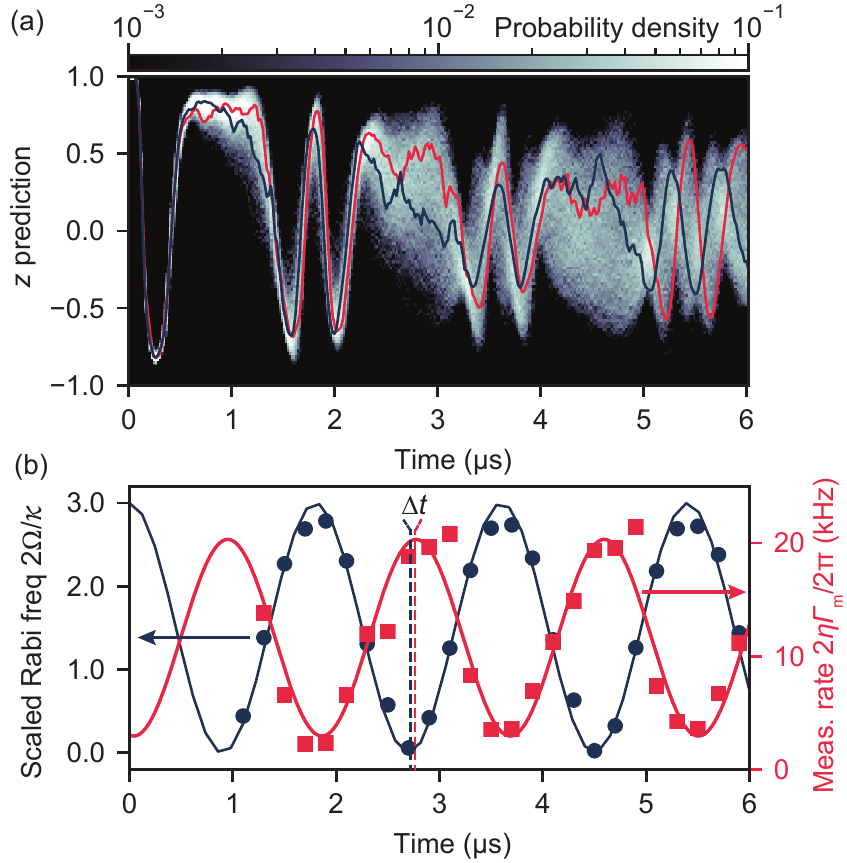}
    \caption{Uncovering hidden time-dependencies from trajectories subject to sinusoidal Rabi drive modulation. (a) A histogram of the reconstructed $z$ coordinate shows periods of increased and decreased diffusion, also visible in two sample trajectories (red and blue). (b) Repeated analysis of the LSTM trajectories for successive 0.2~\textmu{}s-long time windows. For each window we extract the instantaneous Rabi frequency and measurement rate by fitting the average drift to Eqs.~\eqref{eq:avg_drift_dy}-\eqref{eq:avg_drift_dz} and  diffusion to Eqs.~\eqref{eq:diffusion_dy}-\eqref{eq:diffusion_dz}, respectively. The extracted measurement rate $2\eta \Gamma_m/2\pi$ (red squares) shows maxima during periods of large diffusion in (a). The solid red line is a simple sinusoidal fit to the red squares. The instantaneous Rabi frequencies (blue dots), obtained from fitting trajectories to the Lindbladian part of the reduced qubit master equation, are consistent with the applied drive amplitude (solid blue line). Fit results for $t < 1$ \textmu{}s are not shown, since early during the evolution  trajectories have not spread sufficiently for accurate fit results.}
    \label{fig:fig5}
\end{figure}

The resulting histogram of LSTM trajectories (Fig.~\ref{fig:fig5}a) shows alternating periods of trajectory bunching, where the trajectories are protected from measurement back-action, and diffusion, where the weak measurement clearly imparts stochastic kicks to the qubit. We unravel this non-trivial dynamics by dividing the LSTM trajectories in 0.2~\textmu{}s-long time windows and analyze the average drift and diffusion as described in the previous section. During each 0.2~\textmu{}s window, the trajectories only partly cover the $yz$-plane, but we can still extract the Rabi frequency from the average drift and measurement rate from the diffusion (Fig~\ref{fig:fig5}b). Without prior knowledge of the time-dependence of $\Omega, \Gamma_d$ or $\Gamma_m$, we correctly recover the sinusoidal applied Rabi frequency (Fig~\ref{fig:fig5}b, blue), and find that it anti-correlates strongly with the extracted measurement rate (Fig~\ref{fig:fig5}b, red), consistent with the results in Fig.~\ref{fig:fig4}c. A sinusoidal fit to the instantaneous measurement rate shows a delay of $\Delta t = 0.14 \pm 0.06$ \textmu{}s with respect to the instantaneous Rabi frequency, consistent with the resonator relaxation time $2/\kappa$. The results of Fig~\ref{fig:fig5} demonstrate that the LSTM correctly reconstructs trajectories with hidden time-dependencies, and that parameters can be extracted with high time resolution, ultimately limited by the time step $\mathrm{d}t$ of the voltage record. 

\section{Conclusions}
We have shown that LSTM recurrent neural networks can outperform standard quantum filters if the qubit dynamics is faster than the resonator linewidth. In this previously unexplored regime of weak measurement, the LSTM correctly recovers Lindbladian dynamics and distorted measurement backaction without prior knowledge of the resonator memory. For the strongest Rabi drives the LSTM validation error remains low; however, in this case the resonator temporally averages most of the qubit signal, reducing the best trajectory estimate to an ensemble-averaged trajectory. Therefore, the LSTM is most useful in the intermediate regime $2\Omega/\kappa < 2$, where the measurement backaction is distorted, but the backaction strength is non-zero and trajectories are therefore non-trivial. We further showed that the LSTM is not constrained to constant Hamiltonians. Even when underlying parameters vary on a sub-microsecond timescale the LSTM accurately recovers quantum trajectories, which allows parameter estimation with high time resolution, and may enable trajectory reconstruction for qubits connected to strong non-Markovian environments \cite{Niu2019Arxiv, Krastanov2020} where coupling rates can vary with time.

To further improve the LSTM accuracy for applications such as parameter estimation, it is possible to add physical constraints to the LSTM loss function \cite{Banchi2018,BeuclerPRL2021,Genois2021}. In addition, other neural network architectures such as tensor networks or models based on dilated causal convolutions \cite{Oord2016Wavenet} may improve prediction accuracy, though such comparisons require more training data. Finally, a possible extension of our LSTM is state tracking in larger Hilbert spaces. This requires more projective measurement pulses (e.g. at least 4 for single qutrits or 9 for two qubits) compared with the single qubit state tracking, which only requires 3 tomography axes. Given the single qubit data acquisition time of approximately three hours, extending our LSTM tracking to single qutrits or two qubits seems feasible, though optimizing data transfer from the digitizer to computer memory may be desirable to keep data acquisition efficient.

The authors would like to thank K. Nowrouzi, B. Mitchell and A. Morvan for experimental assistance, and L.~B.~Nguyen and Y.~Kim for valuable comments on the manuscript. We gratefully acknowledge insightful conversations with L. Martin, \'E. Genois, A. DiPaolo, J. Gross and M. Laskin. This work was supported by the U.S. Army Research Laboratory and the U.S. Army Research Office under contract/grant number W911NF-17-S-0008. L.~B., S.~G. and J.~D. acknowledge additional support from NSF-BSF Grant Award No.~1915015.

\appendix
\titleformat{\section}{\bfseries\normalsize}{APPENDIX \thesection:}{1em}{\MakeUppercase}
\renewcommand{\thesection}{\Alph{section}}
\renewcommand{\thesubsection}{\arabic{subsection}}
\setcounter{equation}{0}


\section{Experimental setup}
\label{chap:experimental_setup}

\begin{figure*}[t]
    \centering
    \includegraphics[width=\textwidth]{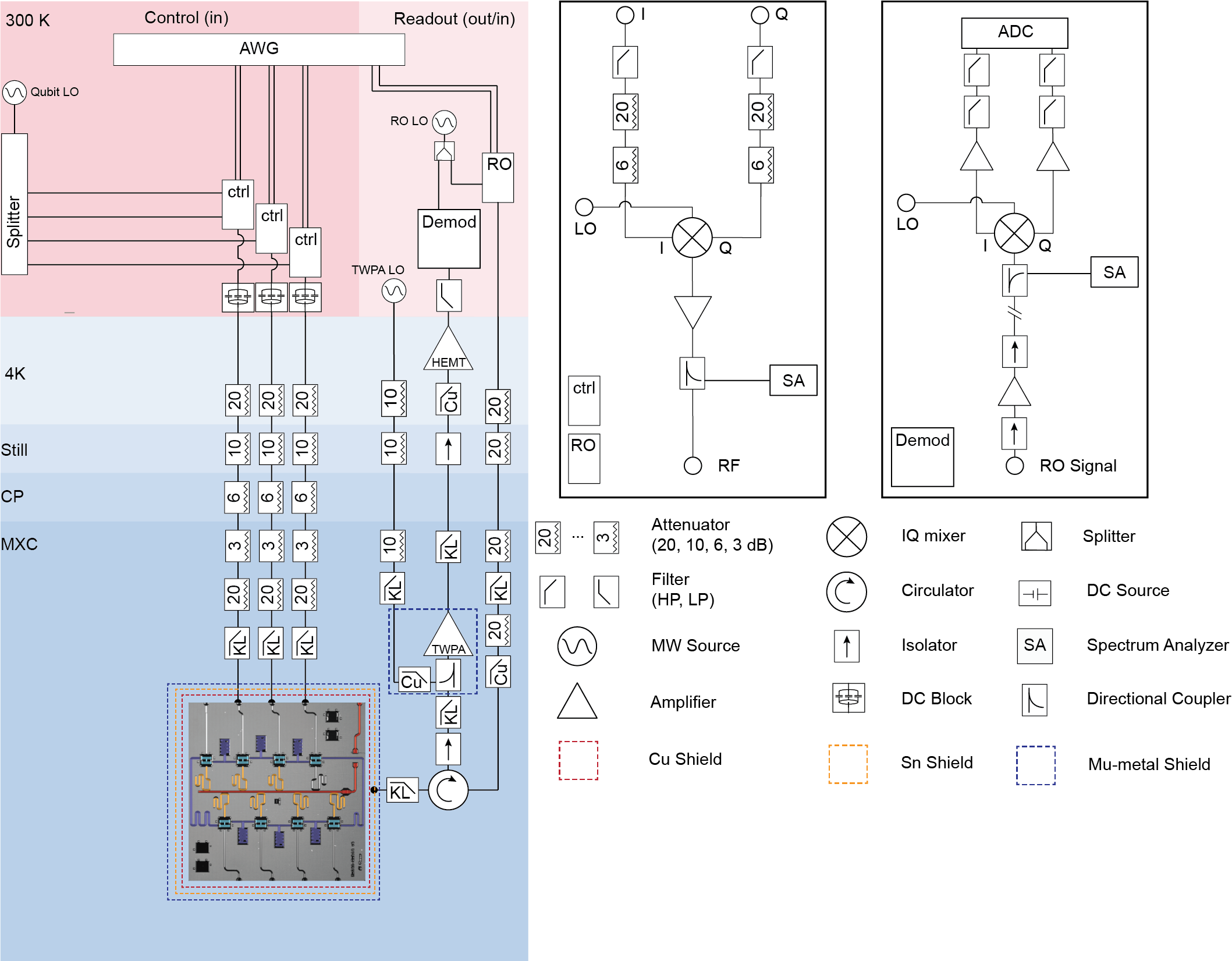}
    \caption{Microwave electronics inside and outside the dilution refrigerator. CP = cold plate (100 mK), MXC = mixing chamber (10 mK). Any additional description of terms can be found in Appendix \ref{chap:experimental_setup}.}
    \label{fig:experimental_setup}
\end{figure*}

The experiments in this work were performed with a chip of 8 superconducting transmon qubits, cooled to 10 mK in a BlueFors XLD dilution refrigerator. Room-temperature and cryogenic electronics for qubit control and measurement are shown in Fig. \ref{fig:experimental_setup}. Qubit control pulses are generated by upcoversion of intermediate frequency (IF) pulses generated by a Keysight PXI arbitrary waveform generator (AWG) via in-phase quadrature (IQ) modulation of a continuous wave (CW) local oscillator (LO) tone, sourced by a Keysight MXG N5183B at 5.415 GHz. Both I and Q components are sourced at 1 GS/s between 66 and 261 MHz. The phase and DC offsets between the I and Q waveforms are tuned to eliminate the opposite sideband and LO leakage due to mixer nonidealities, while band-pass filtering at room-temperature reduces noise from the AWG. Readout pulses are generated with the same AWG and are similarly upconverted with a 6.83 GHz LO tone from a separate Keysight MXG N5183B. Control pulses are passed through a DC block, and all pulses travelling to the sample are attenuated at each stage, with a further K\&L low-pass filter at the base stage. The upconversion chain schematic for both the qubit control and readout lines is detailed in the inset labeled ``ctrl'' and ``RO''.

Reflected measurement signals are redirected by a circulator to a measurement chain outfitted with superconducting coaxial cable, where they are amplified by a traveling wave parametric amplifier (TWPA) at 10 mK, a high electron mobility transistor (HEMT) at 4K, and a low noise room temperature amplifier, before being downcoverted to IQ IF components with the same 6.83 GHz LO tone used for RO upconversion. The downconverted signals are then amplified and filtered to reduce high-frequency amplifier noise before being digitized at 1 GS/s by an Alazar analog-to-digital converter (ADC) and demodulated in software. The downconversion chain schematic is detailed in the inset labeled ``Demod''.

The TWPA is pumped with a CW tone, sourced by a Hittite HMC M2100 at 7.42 GHz and 16.5 dBm before being attenuated and filtered by low-pass K\&L and copper powder filters. 

\section{Device parameters and stability}
\label{chap:device_parameters_and_stability}

\begin{table*}
\caption{Calibrated resonator and qubit parameters for the device used in this work. Note that $\omega_\mathrm{res}$ and $\omega_\mathrm{ge}$ are the bare readout resonator and qubit frequencies, respectively and that $T_1$ and $T_{2, \mathrm{Ramsey}}$ were measured without a weak measurement tone ($\varepsilon/2\pi = 0$). The  error bars on $T_1$ and $T_{2, \mathrm{Ramsey}}$  reflect the standard deviation of 100 repeated measurements during a 3-hour window.}
\label{tab:res_qubit_params}
\begin{tabular}{@{}llc@{}}
\toprule
\multicolumn{1}{c}{Parameter}     & Description & Measured value  \\
\midrule
$\omega_\mathrm{res}/2\pi$       & Bare cavity frequency & 6679 MHz           \\
$\kappa/2\pi$              & Cavity linewidth & 1.56 MHz           \\
$\omega_\mathrm{ge}/2\pi$       & Transmon frequency (empty cavity) & 5473 MHz           \\
$\alpha/2\pi$              & Transmon anharmonicity & -270 MHz           \\
$T_1$                     & Transmon relaxation time ($\Omega = 0$) & 61 $\pm$ 7 \textmu{}s     \\
$T_{2, \mathrm{Ramsey}}$  & Ramsey dephasing time (empty cavity) & 70 $\pm$ 9 \textmu{}s     \\
$\chi / 2\pi$               & Half of the dispersive cavity shift & 0.47 MHz           \\
$\varepsilon/2\pi$           & Weak measurement strength & 0.43 $\pm$ 0.01 MHz \\
\bottomrule
\end{tabular}
\end{table*}

Table \ref{tab:res_qubit_params} summarizes the relevant parameters of the qubit and readout resonator used in this work. We obtain the resonator frequency $\omega_\mathrm{res}$ and cavity linewidth $\kappa/2\pi$ through spectroscopic measurements, and we measure the qubit frequency $\omega_\mathrm{ge}$, the anharmonicity $\alpha/2\pi$ and $T_{2, \mathrm{Ramsey}}$ through repeated Ramsey measurements on the g-e transition ($\omega_\mathrm{ge}$, $T_{2, \mathrm{Ramsey}}$) and e-f transition ($\alpha/2\pi$). Lastly, we fit the weak measurement strength $\varepsilon / 2\pi$ by comparing the measured Rabi decay time as function of Rabi frequency to a master equation simulation using QuTiP \cite{qutip2013}. The errorbar on $\varepsilon$ is due uncertainty in the fit. 

Acquiring a training data set (i.e. one initial qubit state with three tomography axes) for a single Rabi frequency takes approximately three hours. As the neural network learns a representation of the qubit Hamiltonian, it is important that the parameters remain stable during the acquisition.

Since readout infidelity and unexpected decay or dephasing directly affect the training labels, we have characterized the stability of $T_1$, $T_{2, \mathrm{Ramsey}}$ as well as the readout fidelity over a three-hour time span. For $T_1$ and  $T_{2, \mathrm{Ramsey}}$ we observe fluctuations on the order of 10-15\% of their respective values, which is likely due to nearby two-level systems that couple weakly to the qubit. Since the fluctuations are modest, and average values of 61 and 70 \textmu{}s remain much larger than the maximum trajectory length of 8.0 \textmu{}s, we do not expect decay time fluctuations to affect the performance of the neural network.

We characterize readout infidelity by repeatedly preparing the qubit in the ground and excited state followed by a measurement of the qubit state. We only observe small fluctuations of the correctly classified state over a three-hour window and find average correctly classified probabilities $P(0|0) = 0.995 \pm 0.005$ and $P(1|1) = 0.984 \pm 0.004$. We expect that the average readout infidelity of $1 - \frac{1}{2} (P(0|0) + P(1|1)) = 1.0\%$ affects the accuracy of the trained neural network, but fluctuations in this quantity do not play a significant role. 

The qubit studied in this work is part of a larger 8-qubit quantum processor, and is capacitively coupled to two neighboring qubits, each via a coupling bus with an effective coupling strength of approximately 2 MHz. To avoid any undesired qubit dynamics due to the neighboring qubits, we post-select the  voltage records, conditioning on all qubits starting in the ground state before the Rabi sequence, and the neighboring qubits remaining in the ground state after the Rabi sequence. Typically the post-selection removes 5\% of the voltage records.

\section{Processing of measurement records}
Traditional processing of weak measurement records includes digital filtering and choosing a coarse grained $\mathrm{d}t$ that minimizes temporal correlations in the measurement record, to ensure that the appropriate stochastic master equation can be applied faithfully to the measured data \cite{Campagne-IbarcqPRX2016, Ficheux2018}. As explained in the main text, for fast qubit dynamics it becomes impossible to coarse grain the record in a way that produces uncorrelated measurement records without undersampling the data. We strike a balance between undersampling and correlation by coarse graining the measurement records as listed in Table~\ref{tab:supp_dataset_props}. Note that all records are sampled faster than the uncorrelated condition $\mathrm{d}t > 2 / \kappa = 0.2$~\textmu{}s, and the two fastest datasets are sampled more finely than all others ($\mathrm{d}t = 20$~ns). Feeding the Bayesian quantum filter correlated measurement records is expected to contribute to the RMS error in Fig.~\ref{fig:fig2}d. Interestingly, the Bayesian filter remains surprisingly accurate up to $2\Omega/\kappa \approx 1$. 

\begin{table*}
\caption{Properties of the data sets used for Figs.~\ref{fig:fig2}-\ref{fig:fig4} in the main text. The column header ``Init qubit state'' refers to the prepared initial qubit state. ``Data set size'' is the total number of trajectories available for each data set, which includes a fraction of trajectories used for analysis (``Traj. for LSTM analysis''). The subsequent columns describe the longest Rabi drive duration ($T_m$), the sampling increment of the Rabi drive duration, and the coarse graining of the voltage records. Finally, ``Sequence length'' refers to the number of samples in each voltage record, which is given by the ratio of the Longest $T_m$ and $\mathrm{d}t$. Note that all voltage records within a data set have equal sequence length, since sequences with Rabi drive durations smaller than $T_m$ are zero-padded.}
\label{tab:supp_dataset_props}
\begin{tabular}{cccccccc}
\toprule
$2\Omega/\kappa$ & Init qubit state & Data set size     & Traj. for LSTM analysis & Longest $T_m$ (\textmu{}s) &  $T_m$ increment (ns) & $\mathrm{d}t$ (ns) & Sequence length \\
\midrule
0.0              & $|+Y\rangle$        & $454\times10^3$   & 46,579                         & 8.0                                         & 200                          & 40                                & 200                 \\
0.0              & $|-Y\rangle$        & $454\times10^3$   & 46,540                         & 8.0                                         & 200                          & 40                                & 200                 \\
0.2              & $|+Z\rangle$        & $283 \times 10^3$ & 29,027                         & 8.0                                         & 200                          & 40                                & 200                 \\
0.6              & $|+Z\rangle$        & $283 \times 10^3$ & 29,029                         & 8.0                                         & 200                          & 40                                & 200                 \\
1.0              & $|+Z\rangle$        & $283 \times 10^3$ & 29,014                         & 8.0                                         & 200                          & 40                                & 200                 \\
1.4              & $|+Z\rangle$         & $283 \times 10^3$ & 29,062                         & 8.0                                         & 200                          & 40                                & 200                 \\
2.0              & $|+Z\rangle$         & $353 \times 10^3$ & 36,054                         & 5.0                                         & 100                          & 20                                & 250                 \\
3.2              & $|+Z\rangle$         & $354\times10^3$   & 36,118                         & 5.0                                         & 100                          & 20                                & 250                \\
\bottomrule
\end{tabular}
\end{table*}

\section{Training of the neural network}
\label{chap:network_training}
\subsection{Training data acquisition details}
The pulse sequence for training data acquisition is given in Fig. \ref{fig:supp_convergence}a. For each experimental run, the qubit is heralded in the ground state $\ket{0}$ with a strong measurement, resulting in a large mean photon occupation number $\bar{n}$ in the cavity. The measurement tone amplitude is then reduced to the weak measurement setpoint ($\bar{n} \approx 0.3$), and a 5 \textmu{}s idling time allows $\bar{n}$ to reach steady state. Next, we prepare the qubit in one of the cardinal points of the Bloch sphere with a 30 ns preparation pulse, which is detuned from $\omega_{ge}$ to account for the Stark shift \cite{Schuster2005} induced by the non-zero cavity $\bar{n}$. At $t=0$ we introduce $\sigma_x$ dynamics by quickly ramping up the Rabi drive (ramp time 5 ns), while continuing to weakly measure the qubit along $\sigma_z$. This yields voltage records ($I_n, Q_n$) of variable duration $T_m$, which ranges between 0 and 8 \textmu{}s. Finally, a projective measurement along one of the principal axes of the Bloch sphere $\sigma_{x, y, z}$ is performed by first applying a single qubit gate to rotate the qubit and then a strong measurement ($\bar{n} \gg 1$). The projective measurement results are binary variables $P_i \in \{0, 1\}$ and represent the training labels for the neural network.

\subsection{Training details}

\begin{figure*}
    \centering
    \includegraphics[width=\textwidth]{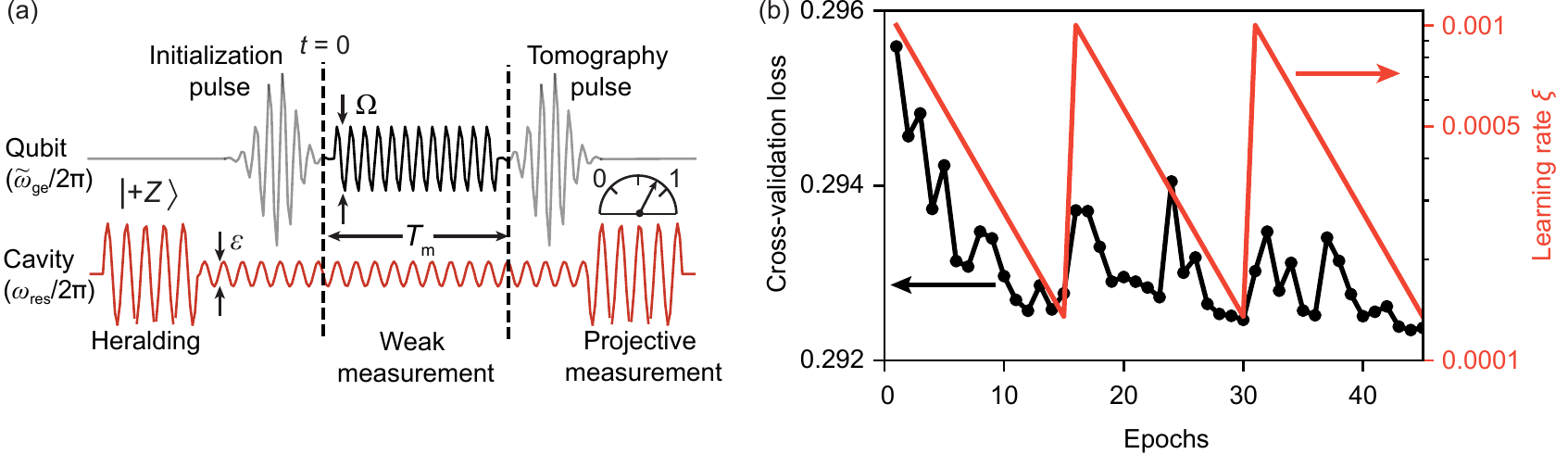}
    \caption{(a) Experimental pulse sequence for training data acquisition. Time axis is not to scale. (b) Cost function (black, left axis) computed on the cross-validation data set during a typical training session. Temporary increases in the cost function are expected as the learning rate (red, right axis) increases periodically.}
    \label{fig:supp_convergence}
\end{figure*}

In this work we use a long short-term memory recurrent neural network, a model that recurrently updates a layer of virtual neuron-like nodes as it processes a time-serialized input, and features a memory cell to retain past information of the time series \cite{Hochreiter1997, Gers2000}. At time $t$, the layer is encoded in a vector $\vec{h}_t$, and propagates to time $t+1$.

A typical data set with single initial qubit state, and a single Rabi frequency consists of $N \approx 4 \times 10^3$ repetitions of $M = 40$ different length voltage records, repeated for three tomography axes. We use 90\% of the approximately $3 M N = 4 \times 10^5$ voltage records for training and the remaining 10\% for cross-validation (see Table~\ref{tab:supp_dataset_props}).

The training code uses the Tensorflow 2 library in Python and consists of several epochs where training samples are fed forward and backwards through the neural network in mini-batches of size $N_b=512$. During the training weights and biases of the network are adjusted to minimize a cost function $\mathcal{L}$, which contains the following three components (similar to Ref.~\cite{Genois2021}): 
\begin{itemize}
    \item A cross-entropy loss associated with the projective measurement result at the end of a voltage record $$\mathcal{L}_\mathrm{CE} = - \frac{1}{N_b} \sum_{i=1}^{N_b} P_i \log (s_i) + (1-P_i) \log(1-s_i),$$ where $P_i$ and $s_i$ are the tomography result (0 or 1) and probability assigned by neural network ($0 \leq s_i \leq 1$) for voltage record $i$ in the mini-batch, respectively.
    \item A mean-squared-error for deviating from the known initial state at time $t = 0$, $\mathcal{L}_\mathrm{init}$. 
    \item A Rectified Linear Unit (ReLU) activation on the purity of the quantum state $\mathcal{L}_\mathrm{purity}$ at all times, which tries to enforce that the Bloch vector does not lie outside the Bloch sphere for all $t$. 
\end{itemize}
We construct the loss function as a weighted sum of the components above, but note that the main contribution to the loss function comes from $\mathcal{L}_\mathrm{CE}$. 

We use the \textsc{adam} optimizer \cite{ADAM2014} to obtain the LSTM parameters that minimize the cost function $\mathcal{L}$, and adjust the step size of the stochastic gradient descent (i.e. learning rate) $\xi$ in a cyclical fashion (Fig \ref{fig:supp_convergence}b) to prevent getting stuck in local minima \cite{SmithCyclicalLR2017}. We have experimented with adding dropout and regularization, techniques which aim to improve generalization on unseen data \cite{SrivastavaDropoutJLMR2014}, but have not observed improvements in $\mathcal{L}$ computed on the validation dataset. Since dropout randomly resets a fraction of trainable parameters between epochs, it destroys temporal correlations between successive LSTM cells which may be hard to learn from the training data. In addition, we believe that the lack of improvement could be due to the noisy nature of the voltage records and to the stochastic nature of the projective measurements (training labels), which could prevent overfitting even in the absence of dropout and regularization.

During the training process we monitor the value of $\mathcal{L}$ for both the training and cross-validation samples (for an example see Fig \ref{fig:supp_convergence}b). The neural network is converged when $\mathcal{L}$ computed on the training dataset no longer decreases after an annealing cycle \cite{Graves2013}. The cross-validation samples are then passed through the network once more and we use the resulting trajectories in Figures 2 through 5 of the main text. For our largest dataset of $0.8 \times 10^5$ voltage records, the entire training process completes within approximately 20 minutes thanks to the processing power of the \textsc{gpu} (NVidia GeForce RTX 2080 Ti). Finally, the code used for training is available on Github \cite{TrajectoriesLSTMGitHubRepo}. A link to the training data will be available here later.

\section{Trajectory histograms}
\label{chap:traj_histograms}
The trajectory histograms shown in Fig. \ref{fig:fig2}a of the main text show that for $t \gg \Gamma_d^{-1}$ trajectories never cross a threshold in $z$, and therefore seem constrained to a reduced-radius Bloch sphere. In this section we derive this effect from the Stochastic master equation. 

We consider the Stochastic master equation (in It\^o form) for heterodyne measurement of a driven qubit, where the Rabi drive is applied in the $yz$ plane \cite{AtalayaPRL2019}: 
\begin{align}
  &\mathrm{d}x = -\Gamma_d x \, \mathrm{d}t  - \sqrt{2 \eta \Gamma_m} z x \, \mathrm{d} W_1 - \sqrt{2 \eta \Gamma_m} y \, \mathrm{d} W_2 \label{eq:sme_het_dx}\\
  &\mathrm{d}y = -\Gamma_d y \, \mathrm{d}t + \Omega z \, \mathrm{d}t - \sqrt{2 \eta \Gamma_m} z y \,  \mathrm{d} W_1 + \sqrt{2 \eta \Gamma_m} x \, \mathrm{d} W_2 \label{eq:sme_het_dy}\\ 
  &\mathrm{d}z = -\Omega y \, \mathrm{d}t + \sqrt{2 \eta \Gamma_m} (1 - z^2) \, \mathrm{d} W_1.\label{eq:sme_het_dz}
\end{align}
These equations include informational backaction ($\mathrm{d} W_1$) causing trajectories to collapse to the poles $\ket{\pm Z}$ at a rate $\eta \Gamma_m$, phase backaction ($\mathrm{d} W_2$) at an equal rate due to the quantum fluctuations in the cavity photon number $\bar{n}$, and dephasing towards the origin of the Bloch sphere at a rate $2(1-\eta)\Gamma_m$. Note that we define $\Gamma_m$ as the decoherence rate due to measurement of a single quadrature, and since we measure both $I$ and $Q$ quadratures the total dephasing rate includes a factor of 2. Further note that for a unit efficiency measurement setup, individual trajectories diffuse quickly but remain pure. Therefore, the reduced radius is due to limited measurement efficiency $\eta$.

To further quantify the reduced radius observed in the experimental trajectories, we calculate the mean change of the purity, defined as $|\mathbf{r}|^2 = x^2 + y^2 + z^2$:
\begin{align}
    \left \langle \frac{\mathrm{d}|\mathbf{r}|^2}{\mathrm{d}t} \right \rangle = &-2 \left[ (1-\eta)\Gamma_m + \Gamma_m \right] (|\mathbf{r}|^2 - z^2) \notag \\
    &+ 2 \eta \Gamma_m(1-2z^2) + 2 \eta \Gamma_m |\mathbf{r}|^2 z^2
\end{align}
We can simplify this expression in the fast Rabi regime ($\Omega \gg \Gamma_m$), where $z^2 \approx |\mathbf{r}|^2 / 2$ evaluated over one Rabi cycle. This yields
\begin{equation}
    \left \langle \frac{\mathrm{d}|\mathbf{r}|^2}{\mathrm{d}t} \right \rangle = -2 \eta \Gamma_m - (2 + \eta) \Gamma_m |\mathbf{r}|^2 + \eta \Gamma_m |\mathbf{r}|^4 \label{eq:supp_r_sq_fast}.
\end{equation}

\begin{figure}
    \centering
    \includegraphics[width=0.4\textwidth]{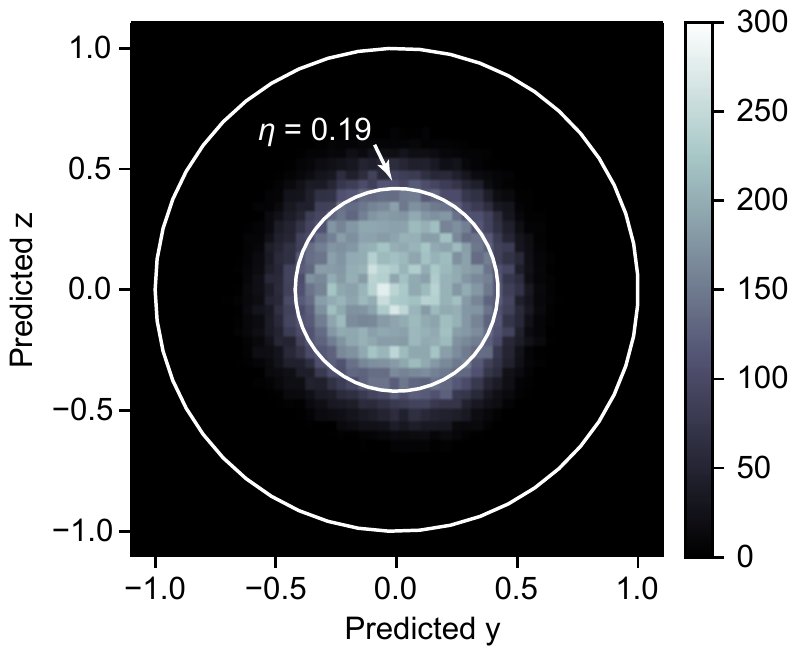}
    \caption{Trajectory histogram for a fast Rabi drive $2\Omega/\kappa = 2.0$, post-selected for steady state (times longer than $\Gamma_d^{-1}$). The inner circle is a prediction for the purity of individual trajectories from Eq. \eqref{eq:supp_st_state_radius} with $\eta$ calibrated from Appendix \ref{chap:efficiency_calibration}.}
    \label{fig:supp_driven_yz_hist}
\end{figure}

From Eq.~\eqref{eq:supp_r_sq_fast} we calculate the steady state solution for the reduced radius
\begin{equation}
    |\mathbf{r}| = \sqrt{\frac{2 + \eta}{2\eta} - \sqrt{\left(\frac{2 + \eta}{2\eta} \right)^2 - 2}}.
    \label{eq:supp_st_state_radius}
\end{equation}
which depends entirely on the measurement efficiency, and requires a small correction if the environmental dephasing rate is nonzero. Using the efficiency calibration from Appendix \ref{chap:efficiency_calibration}, we find that the predicted radius $|\mathbf{r}|$ agrees well with the observed trajectories histogram shown in Fig. \ref{fig:supp_driven_yz_hist} for $2\Omega/\kappa = 2.0$. While the same physics applies to the histogram in Fig. \ref{fig:fig2} of the main text, the predicted $|\mathbf{r}|$ is inaccurate for this moderate Rabi rate $\Omega/\Gamma_m \approx 1$, since the approximation $\Omega \gg \Gamma_m$ is invalid. In this regime further corrections to Eq. \eqref{eq:supp_r_sq_fast} are necessary, but these are beyond the scope of this work.

\section{Accuracy of the predicted trajectories by the LSTM}
\label{chap:rnn_accuracy}
In Fig \ref{fig:fig2}b of the main text, we demonstrate the accuracy of the LSTM trajectories for a slow Rabi drive $2\Omega/\kappa = 0.2$. The purpose of this appendix is to define the accuracy and to discuss error sources. 

To quantify the LSTM accuracy, we bin all predictions immediately before a projective measurement associated with each trajectory and average the projective measurement results for trajectories with similar predictions. This technique is frequently used to experimentally verify an estimate of the qubit state with an independent measurement \cite{Murch2013,Weber2014}. We use the following definition of the weighted root mean square (RMS) error for the $x$ coordinate ($y$ and $z$ are similar): 
\begin{equation}
    \varepsilon_x = \sqrt{\sum_{i = 1}^{N_x} N_i (\underbrace{x_{\mathrm{LSTM}, i}}_\text{LSTM prediction} -  \underbrace{\langle x_{\mathrm{proj}, i} \rangle}_\text{Avg. tomography} )^2  / \sum_{i=1}^{N_x} N_i}
    \label{eq:supp_epsilon_definition}
\end{equation}
where $N_x$ is the number of bins, $x_{\mathrm{LSTM}, i} \in [-1, 1]$ is the predicted value by the LSTM for bin $i$, and $\langle x_{\mathrm{proj}, i} \rangle$ is the average of projective measurements for predictions within $\delta = 2 / N_x$ of $x_{\mathrm{LSTM}, i}$. The definition of Eq. \eqref{eq:supp_epsilon_definition} captures any inaccuracy caused by the LSTM (e.g. inefficient model, insufficient training data) but is also sensitive to the uncertainty in our best estimate of the ground truth $\langle x_{\mathrm{proj}, i} \rangle$, which scales as $1 / \sqrt{N_i}$. This becomes a problem for rare predictions, when the number of trajectories $N_i$ is small. Therefore, by weighing each bin by the number of predictions in that bin $N_i$, $\varepsilon_x$ becomes less sensitive to limited statistics for rare predictions. 

\begin{figure*}
    \centering
    \includegraphics[width=\textwidth]{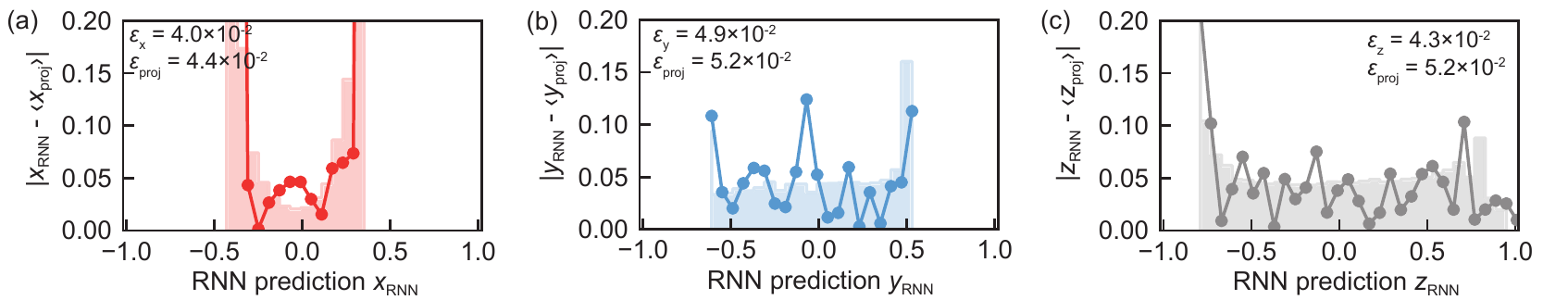}
    \caption{Comparison of the LSTM prediction error (dots) and expected uncertainty in the estimate of the ground truth $\langle x_{\mathrm{proj}, i}\rangle$ (histograms calculated from Eq. \eqref{eq:supp_uncertainty_estimate}) for Bloch coordinates $x$ (a), $y$ (b), and $z$ (c). $\varepsilon_{x,y,z}$ are the weighted RMS errors calculated from LSTM predictions, and $\varepsilon_\mathrm{proj}$ is calculated from Eq. \eqref{eq:supp_epsilon_definition} while setting $x_{\mathrm{LSTM}, i} - \langle x_{\mathrm{proj}, i} \rangle = \sigma_{\langle x_{\mathrm{proj}, i} \rangle}$. $\varepsilon_\mathrm{proj}$ thus represents the weighted RMS error in case of perfect LSTM predictions and 1$\sigma$ statistical fluctuation in $\langle x_{\mathrm{proj}, i} \rangle$ due to a limited number of trajectories. Since both quantities follow similar trends and their magnitudes $\varepsilon_{x,y,z}$ and $\varepsilon_\mathrm{proj}$ are consistent we conclude that our accuracy measure is limited by the number of trajectories.}
    \label{fig:rnn_accuracy}
\end{figure*}

In Fig \ref{fig:rnn_accuracy} we show that the uncertainty in the projective measurement dominates any errors caused by the LSTM, since the absolute error $| (x_{\mathrm{LSTM}, i} - \langle x_{\mathrm{proj}, i} \rangle) |$ closely follows the expected uncertainty in $\langle x_{\mathrm{proj}, i} \rangle$, estimated from the Bernoulli distribution:
\begin{equation}
    \sigma_{\langle x_{\mathrm{proj}, i} \rangle} = \sqrt{\frac{1 - \langle x_{\mathrm{proj}, i} \rangle ^2}{N_i}}.
    \label{eq:supp_uncertainty_estimate}
\end{equation}
We observe similar trends in the absolute error and the uncertainty estimated from Eq. \eqref{eq:supp_uncertainty_estimate} for all Bloch coordinates (Fig \ref{fig:rnn_accuracy}a-c), and error magnitudes are consistent. In other words, our ability to quantify the LSTM error is limited by our knowledge of the true quantum state. To further assess the LSTM performance, more trajectories are needed to reduce our uncertainty in the estimate of the true quantum state. Alternatively, it is possible to train the LSTM on simulated trajectories for which the true quantum state is known \cite{Genois2021}.

\section{Cavity corrections to the Bayesian filter}
\label{chap:bsu_corrections}

Quantum trajectory theory describes how an observer's estimate of a quantum state evolves as it is updated with a weak measurement record \cite{Weber2014, GambettaTrajectories2008}. The measurement record is translated to quantum state evolution by first applying the unitary evolution and then updating the state with the measurement record at each time step, as the backaction of the measurement on the state can be derived \cite{Hacohen-Gourgy_and_Martin}. Thus, with knowledge of the initial state and the Hamiltonian driving unitary evolution, the density matrix $\rho_{ij} (t_i)$ can be repeatedly updated. 

We follow the procedure outlined in Refs. \cite{Korotkov2016,WeberReview2016}. For each time $t_i$ in the voltage record, we first apply the Lindbladian dynamics 
\begin{equation}
    \rho'(t_i) = \exp (i \lambda dt) \, \rho(t_i),
\end{equation}
where $\lambda$ is a $4 \times 4$ matrix and $\rho(t_i) = (\rho_{00}, \rho_{01}, \rho_{10}, \rho_{11})$. Including a Rabi drive detuning $\Delta$ and a Rabi drive misalignment $\Omega_y$, $\lambda$ takes the form 
\begin{align}
    &\lambda = \begin{pmatrix} 0 & \frac{\Omega_+}{2} & -\frac{\Omega_-}{2} & -i \Gamma_1 \\
    \frac{\Omega_-}{2} & -\Delta & 0 & -\frac{\Omega_-}{2} \\
    -\frac{\Omega_+}{2} & 0 & \Delta & \frac{\Omega_+}{2} \\
    0 & -\frac{\Omega_+}{2} & \frac{\Omega_-}{2} & i\Gamma_1
    \end{pmatrix} \\
    &\Omega_\pm = \Omega_x \pm i\Omega_y \\ 
    & \Delta = \omega_\mathrm{ge} - \omega_\mathrm{rabi} \approx 0 \\
    & \Gamma_1 = \frac{1}{T_1}
\end{align}
Note that $\lambda$ describes the same Lindbladian dynamics that leads to Eqs.~\eqref{eq:sme_het_dx}-\eqref{eq:sme_het_dz} provided $\Omega_y = 0$, and $\mathrm{d}t = t_{i+1} - t_i$ is the time increment of the voltage record. The parameters $\Omega_{x,y}$, $\Delta$ and $\Gamma_1$ are obtained from a fit to the projective measurement results. Next, we update the density matrix taking into account measurement record $(I(t_i), Q(t_i))$ and the partially updated density matrix $\rho'(t_i)$:
\begin{align}
    &\rho_{11} (t_{i+1}) = \frac{\rho_{11}' (t_i) / \rho_{00}' (t_i) \, e^{\alpha_i}}{1 + \rho_{11}' (t_i) / \rho_{00}' (t_i) \, e^{\alpha_i}}  \label{eq:state_update_rot_1}\\
    &\rho_{00} (t_{i+1}) = 1 - \rho_{11}(t_{i+1})  \label{eq:state_update_rot_2}\\
    &\rho_{10} (t_{i+1}) = \rho_{10}' (t_i)  \sqrt{\frac{\rho_{11}(t_{i+1})\rho_{00}(t_{i+1})}{\rho_{11}'(t_i) \rho_{00}'(t_i)}} \, e^{-i\beta_i} \, e^{-2(1-\eta)\Gamma_m dt}  \label{eq:state_update_rot_3}\\
    &\rho_{01} (t_{i+1}) = \rho_{10}^\dagger (t_{i+1}) \label{eq:supp_state_update_rot_4}
\end{align}
where $\alpha_i = \tilde{I}_i \Delta I / \sigma^2$, $\beta_i  = \tilde{Q}_i \Delta I / 2\sigma^2$, $\sigma^2$ is the variance of the noise in the voltage record and finally 
\begin{align}
    &\tilde{I}_i =  I(t_i) - \frac{(I_0 + I_1)}{2} \\
    &\tilde{Q}_i = Q(t_i) - Q_0 \\
    &\Delta I = I_1 - I_0.
\end{align}
$I_0$ and $I_1$ are the steady state coherent state amplitudes conditioned on the qubit in the ground state and excited state, respectively. These state update equations are exact in the absence of a Rabi drive, and if the qubit decay rates are small compared with $\kappa$ \cite{Korotkov2016}. Fig \ref{fig:fig2}d in the main text shows that these equations accurately reconstruct trajectories up to $2\Omega / \kappa \approx 1$ even though they do not contain the cavity. 

\begin{figure*}
    \centering
    \includegraphics[width=\textwidth]{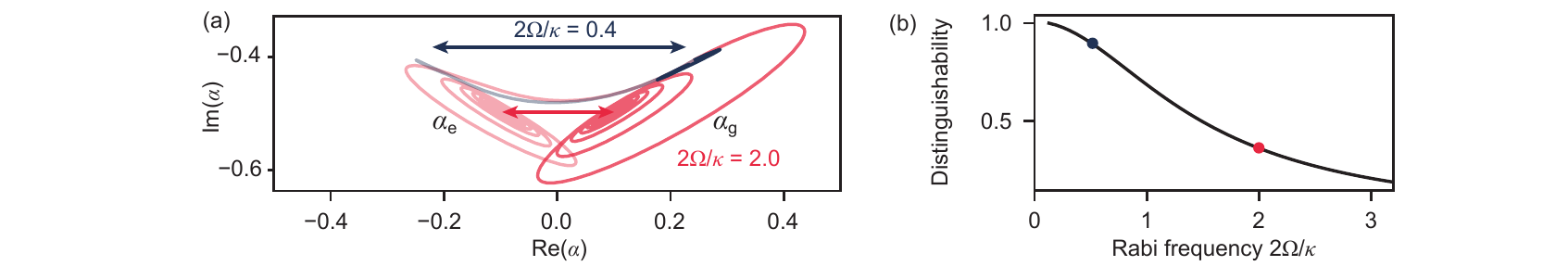}
    \caption{(a) QuTiP master equation simulation of a driven qubit dispersively coupled to the readout cavity, showing coherent state amplitudes vs. time, conditioned on the qubit state. We show two pairs of $\alpha_g$ and $\alpha_e$ for $2\Omega/\kappa = 0.4$ and $2 \Omega / \kappa = 2.0$ obtained from the expectation value $\alpha_g \approx \mathrm{Tr} \left(\rho_\mathrm{q+c} a \ket{g} \bra{g} / P_g \right)$, where $a$ is the cavity photon annihilation operator and $\rho_\mathrm{q+c}$ is the joint qubit-cavity density matrix. (b) We use the relative distinguishability, defined as $\alpha_g(\Omega) - \alpha_e(\Omega) / \alpha_g(0) - \alpha_e(0)$, to scale $\Delta I$ and partially correct for cavity effects in the Bayesian filter method.}
    \label{fig:supp_correction}
\end{figure*}

While $\Delta I$ can be obtained with a simple calibration measurement when $\Omega = 0$, it is impossible to calibrate $\Delta I$ in the presence of a Rabi drive, since the measurement term ($\sigma_z$) and Rabi drive ($\sigma_x$) do not commute. One might assume that $\Delta I$ does not change with the applied Rabi frequency, but this leads to a relatively large RMS error even at moderate Rabi drives. We can improve the performance of the Bayesian filter reconstruction at moderate Rabi drive by adjusting $\Delta I$ based on prior information obtained from a QuTiP numerical simulation which includes qubit and coupling to a cavity. From these simulations (Fig. \ref{fig:supp_correction}a) we find that the steady state coherent state amplitudes conditioned on the qubit state ($\alpha_{g, e}$) substantially decrease as the Rabi frequency increases. Therefore, to correct for cavity effects, we scale $\Delta I$ by the steady state value of $\alpha_e(\Omega) - \alpha_g(\Omega) / (\alpha_e (0) - \alpha_g (0))$ (Fig. \ref{fig:supp_correction}b), which yields the triangles in Fig. \ref{fig:fig2}d of the main text with substantially smaller RMS error. 

While the scaling factor for $\Delta I$ partially corrects for the cavity memory effect, the RMS error remains large for $2\Omega/\kappa \gg 1$, since temporal correlations in the voltage record are large. In this case the simple steady-state Bayesian state update equations are no longer valid, and the most straightforward way to further reduce the validation error is by including the full cavity in the stochastic master equation. However, as stated in the main text, this requires precise knowledge of coupling rates to the cavity, which can be hard to calibrate and may depend on the Rabi frequency (similar to $\Delta I$). In addition, it is difficult to experimentally verify whether the cavity modes are tracked accurately, as this would require cavity state tomography. 

In Fig.~\ref{fig:fig2}d we show a further improvement in validation error of the standard Bayesian filter after incorporating the two main effects of the resonator memory in the state update equations (based on Sec. \ref{chap:supp_deriv_back_map}). We label this method as BF+analytics accordingly. In addition to a reduced measurement backaction, which was already included in the BF+numerics method described above, this method also takes into account the rotation of the measurement eigenstates resulting in a much better filter for fast qubit dynamics. This simple correction assumes that the Rabi drive changes the measured observable appreciably over the timescale of the resonator decay, effectively averaging it along a Rabi oscillation segment. This assumption is valid only for faster drives that can outpace the total measurement dephasing rate. Slower drives (with $\Omega < 2 \Gamma_d$) will have their dynamics dominated by the quantum Zeno effect instead, where the measurement rapidly pins the state to an eigenstate mixture and prevents coherent Rabi evolution from occurring \cite{hacohen2018incoherent}. Thus our expectation is that the BF+analytics method should perform less well than the standard steady-state filter for smaller drives within this quantum Zeno regime, which is supported by the data for the two drive strengths, $\Omega/2\pi \approx 0$ and $0.2$~MHz, smaller than the quantum Zeno threshold of $2 \Gamma_d/2\pi \approx 0.4$~MHz.

\section{Verification of heterodyne measurement backaction}
\label{chap:hetdyne_backaction}
\begin{figure*}
    \centering
    \includegraphics[width=0.85\textwidth]{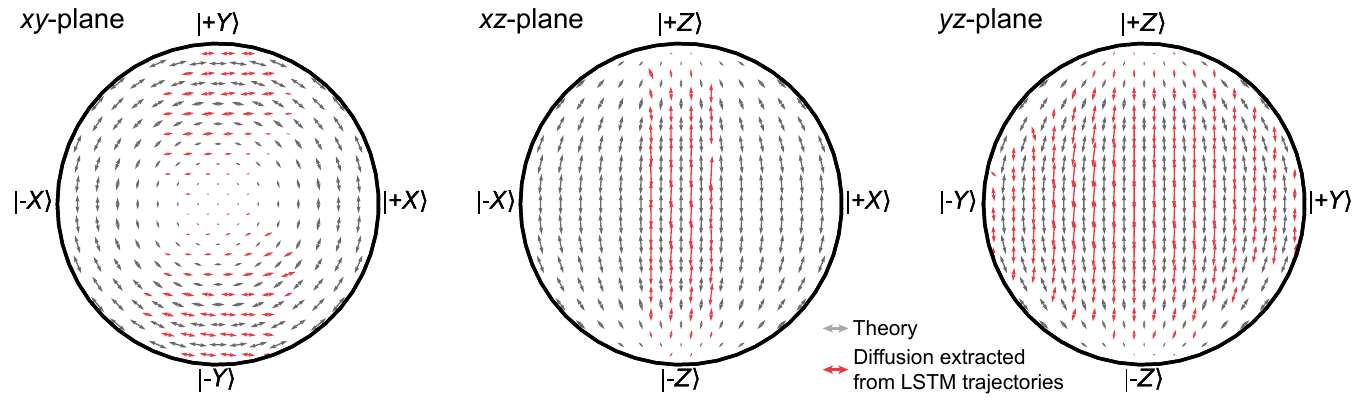}
    \caption{Comparison between theoretical heterodyne measurement backaction (gray quiver maps) and reconstructed measurement backaction by the LSTM (red quiver maps) for an undriven qubit initially prepared in $\ket{\pm Y}$. The theory is obtained by setting $z, y, x = 0$ in the measurement backaction term of Eqs.~\eqref{eq:sme_het_dx}-\eqref{eq:sme_het_dz} for backaction in the $xy$, $xz$ and $yz$ plane, respectively. The reconstructed quiver maps do not cover the full $xy$ and $xz$ plane due to the limited measurement efficiency $\eta$ but agree qualitatively with the theory.}
    \label{fig:hetdyne_backaction}
\end{figure*}

A phase preserving measurement technique gives rise to both informational and phase backaction. In Fig \ref{fig:hetdyne_backaction}, we verify that the undriven trajectories reconstructed by the LSTM indeed show the signatures predicted by Eqs.~\eqref{eq:sme_het_dx}-\eqref{eq:sme_het_dz}. 

We analyze the measurement backaction for dynamics in the $x = 0, y = 0$ and $z = 0$ planes separately and the plots show good agreement with the expected measurement backaction, since we observe phase backaction in the $xy$ plane and informational backaction in the $xz$ and $yz$ planes. This suggests that the LSTM accurately predicts heterodyne measurement backaction from experimental observations, and that the parametrization of the measurement backaction correctly describes the experiment for $2\Omega/\kappa \ll 1$. 

\section{Comparison between voltage records and reconstructed trajectories}
\label{chap:comp_volt_rec_trajs}
\begin{figure*}
    \centering
    \includegraphics[width=.9\textwidth]{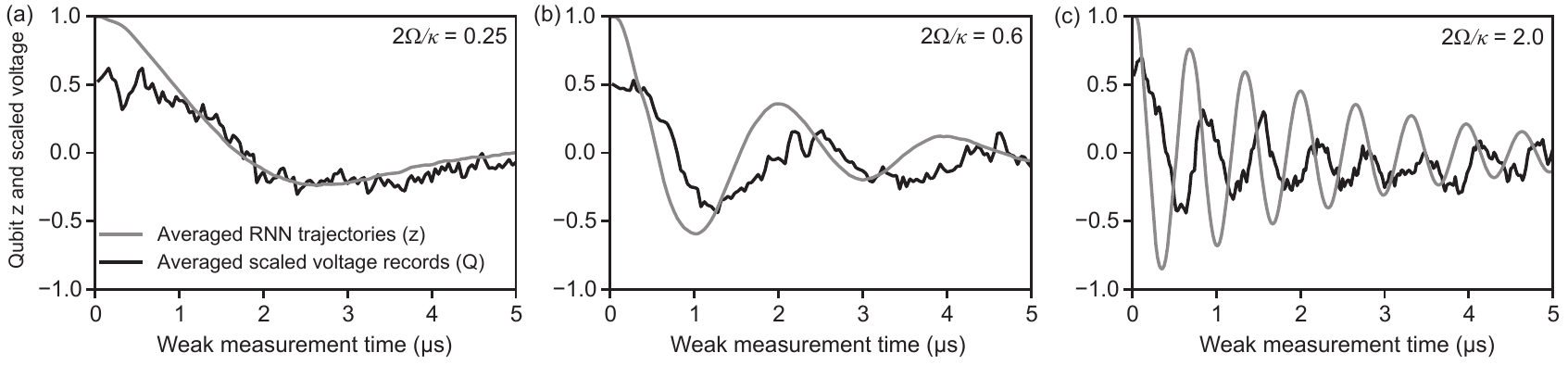}
    \caption{Comparison between averaged voltage records and averaged $z$ trajectories reconstructed by the LSTM for increasing Rabi frequencies (from left to right). Both the trajectories and voltage records are averaged $M=5\times 10^3$ times.}
    \label{fig:supp_comp_volt_rec_trajs}
\end{figure*}

To further understand the breakdown of the Bayesian filter reconstruction demonstrated in Fig \ref{fig:fig2}c of the main text it is helpful to compare the averaged voltage records with the averaged qubit trajectories reconstructed by the LSTM. For fast qubit dynamics (Fig \ref{fig:supp_comp_volt_rec_trajs}c) we observe a large phase difference between the oscillations in the voltage record and those in $z(t)$, because the cavity memory delays photons escaping to the transmission line while the qubit rotates quickly. Conventional stochastic master equations or Bayesian filters that do not include the cavity memory assume photons measure the qubit state $z(t)$ instantaneously and therefore, this phase difference signals the breakdown of those methods. Note that this phase difference already becomes noticeable at intermediate Rabi frequencies (Fig. \ref{fig:supp_comp_volt_rec_trajs}b), consistent with the increase in RMS error at $2\Omega/\kappa = 0.6$ shown in Fig \ref{fig:fig2}d of the main text.

\section{Efficiency calibration}
\label{chap:efficiency_calibration}
\begin{figure*}
    \centering
    \includegraphics[width=\textwidth]{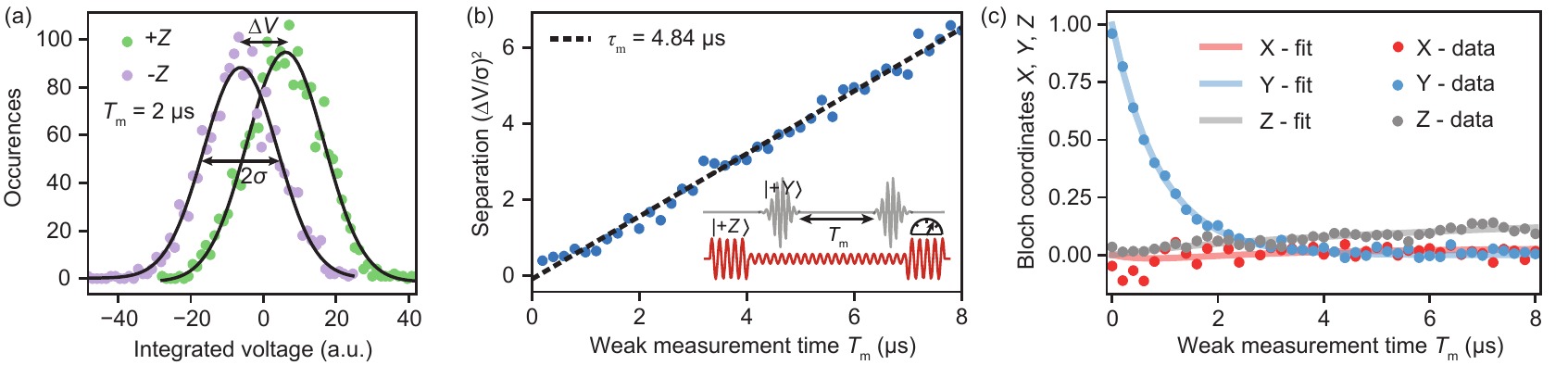}
    \caption{Histograms of the integrated voltage records conditioned on the projective measurement result after preparing the qubit in $\ket{+Y}$ and $T_m = 2.0$ \textmu{}s of weak measurement. The histograms separate by an amount $\Delta V$ due to collapse of the wave function as a result of the weak measurement. (b) The normalized separation $S = (\Delta V / \sigma)^2$ increases linearly with the weak measurement time and the slope gives the characteristic measurement time $\tau_m$. The inset shows a pulse sequence used for this calibration measurement. (c) To extract the efficiency we fit the exponential decay of the projective measurements in the $Y$ basis (blue), which gives $T_d = \Gamma_d^{-1} = 0.91$ \textmu{}s, or $\Gamma_d / 2\pi = 0.175$ MHz.}
    \label{fig:supp_efficiency_calibration}
\end{figure*}

To verify whether the LSTM correctly captures the measurement backaction we compare the extracted measurement rate from the LSTM with an independent calibration measurement, which extracts the separation $S = (\Delta V / \sigma)^2$ of conditional qubit histograms (Fig \ref{fig:supp_efficiency_calibration}a) after weak measurement of variable duration $T_m$. The characteristic measurement time measures how quickly we gain information about the qubit state in the presence of experimental inefficiencies, and is defined by the time it takes to reach a histogram separation of $\Delta V = 2 \sigma$:
\begin{equation}
    \tau_m = 4 \left( \frac{\mathrm{d}S}{\mathrm{d}T_m} \right) ^ {-1}. \label{eq:supp_tau_m}
\end{equation}
In the undriven case where the coherent states in the cavity have reached steady state, the single quadrature measurement rate $\Gamma_m$ is given by \cite{AtalayaPRL2019}
\begin{equation}
    \Gamma_m = \frac{1}{2 \eta \tau_m},
\end{equation}
where $\eta$ is the total efficiency of the measurement chain. In addition, from the definition of the measurement rate we have $\eta = 2\Gamma_m / \Gamma_d$ \cite{FlurinPRX2020}, where the factor of two accounts for both quadratures and $\Gamma_d$ is the decay rate of the ensemble average (Fig \ref{fig:supp_efficiency_calibration}c). The results of the calibration are shown in Fig. \ref{fig:supp_efficiency_calibration}b,c and from Eq. \eqref{eq:supp_tau_m} we find $\tau_m = 4.85 \pm 0.08$ \textmu{}s and $\Gamma_d/2\pi = 0.175$ MHz which gives $\eta = 0.188 \pm 0.003$. This value agrees with the value found from analyzing the stochastic map at $\Omega = 0$ ($\eta = 0.185 \pm 0.002$). Note that this efficiency calibration is not valid when the coherent states of the cavity are not in equilibrium, for example when the qubit is subject to a Rabi drive or when $\Gamma_d > \kappa$. In those cases the methods presented in Ref. \cite{BultinkAPL2018} could offer a way to calibrate the efficiency, but this is beyond the scope of this work.

\section{Derivation of the corrections to the measurement backaction}
\label{chap:tilting_derivation}
\subsection{Problem setup}
A superconducting transmon is measured by coupling it to a strongly detuned readout resonator, which shifts and splits the resonator frequency by a contrast of $2\chi$ between the two qubit-state-conditioned resonances. The ratio $2\chi/\kappa$ of this contrast to the resonance half-width $\kappa/2$ characterizes the distinguishability between resonances, and thus the amount of qubit information per resonator photon that can be stored in correlations. More precisely, given an input drive $\hat{d}(t)e^{-i\omega_d t}$ tuned to the midpoint between qubit resonances, photons in the resonator mode $\hat{a}$ encode this qubit information in their relative phase, which follows from the Heisenberg evolution of the resonator mode in the rotating frame of the drive, $\hat{a}'(t) = -(\kappa/2)[1 + i(2\chi/\kappa)\,\hat{z}(t)]\,\hat{a}(t) + \sqrt{\kappa}\,\hat{d}(t)$. 

\subsection{General derivation}
When the qubit $\hat{z}$ is stationary and the drive $\langle\hat{d}\rangle = -i\varepsilon/\sqrt{\kappa}$ fluctuates around a constant mean, the resulting steady state $\hat{a}^{\rm s.s.} = -i\sqrt{\bar{n}}\,\exp(-i\hat{\phi})$ has a Lorentzian mean photon number $\bar{n} = |2\varepsilon/\kappa|^2/(1 + (2\chi/\kappa)^2)$ and qubit-dependent phase $\hat{\phi} = \arctan[(2\chi/\kappa)\,\hat{z}]$, with a maximum phase contrast of $\Delta \phi_{\rm max} = 2\arctan(2\chi/\kappa)$. A homodyne measurement aligned with the quadrature of maximum separation thus has a maximum amplitude contrast $\Delta \bar{a}_{\rm max} = 2\sqrt{\bar{n}}\sin(\Delta \phi_{\rm max}/2) = \sqrt{\bar{n}}\,(4\chi/\kappa)/\sqrt{1 + (2\chi/\kappa)^2}$ that sets the rate $\gamma_m = (\eta\kappa/2)|\Delta\bar{a}_{\rm max}|^2 = \eta (8\chi^2\bar{n}/\kappa)/(1 + (2\chi/\kappa)^2)$ at which maximally separated steady-states can be distinguished by the photon amplitudes escaping the resonator at rate $\kappa/2$ and being successfully collected with efficiency $\eta$. The uncollected photons and residual qubit-resonator entanglement further contribute to a total qubit ensemble-dephasing rate due to measurement $\Gamma_m \approx \gamma_m/\eta$.

In the presence of a qubit drive, the resonator response additionally filters the evolution of $\hat{z}(t)$ to produce an effectively adiabatic response to a \emph{time-ordered geometric series} of its delay-averages. That is, with a similarly constant drive $\langle \hat{d}\rangle = -i\varepsilon/\sqrt{\kappa}$ and $t\gg 2/\kappa$ to let transients decay, the resonator evolution has the recurrence relation 
\begin{align}\label{eq:qubita}
    \hat{a}(t) &= -i\frac{2\varepsilon}{\kappa} -i\frac{2\chi}{\kappa}\int_0^t\hat{z}(t-\tau)\,\hat{a}(t-\tau)\,\frac{\kappa\, e^{-\kappa\tau/2}\,d\tau}{2},
\end{align}
which implies
\begin{align}
	\hat{a}(t) &= -i\frac{2\varepsilon}{\kappa}\mathcal{T}\sum_{n=0}^\infty\left[-i\frac{2\chi}{\kappa}\int_0^t\hat{z}(t-\tau)\,dP(\tau)\right]^n.
\end{align}

The convolution kernel in the Green's function is an exponential probability distribution $\mathrm{d}P(\tau) = \kappa\,e^{-\kappa\tau/2}\, d\tau/2$ over delay-times $\tau$, normalized as $\int_0^\infty \mathrm{d}P(\tau) = 1$ with mean and variance both equal to the time constant $\tau_c = 2/\kappa$. 

When $\hat{z}$ varies slowly on the timescale $\tau_c$ it can be approximately pulled outside the integral of Eq.~\eqref{eq:qubita} to yield the standard steady-state solution but with a time-dependent phase $\hat{\phi}(t) = \arctan[(2\chi/\kappa)\,\hat{z}(t)]$ that adiabatically tracks the qubit evolution. The next-order approximation treats the evolution as approximately linear within the exponential envelope $\hat{z}(t-\tau) \approx \hat{z}(t) - \hat{z}'(t)\,\tau$, which additionally \emph{delays} the response to the qubit by the mean delay time $\tau_c = 2/\kappa$ to produce the effective phase $\hat{\phi}(t) = \arctan[(2\chi/\kappa)\,\hat{z}(t-\tau_c)]$. For more rapid evolution, part of the evolution is averaged, thus reducing the measurement contrast while rotating the measurement basis.

\subsection{Applying the theory to constant Rabi drive}
In the case of a constant Rabi drive, the delay-average in Eq.~\eqref{eq:qubita} can be computed directly. Assuming dominant harmonic evolution $\dot{\hat{z}} = \Omega\,\hat{y}$, $\dot{\hat{y}} = -\Omega\,\hat{z}$, repeated integration-by-parts of the delay-averaged $\hat{z}(t)$ when $t\gg \tau_c$ yields a pair of geometric series defining an effective $\hat{z}_{\rm eff}(t)$ characterized by an adiabaticity parameter $(2\Omega/\kappa)$,
\begin{align}
   \hat{z}_{\rm eff}(t) &= \int_0^t\hat{z}(t-\tau)\,dP(\tau) \notag \\
   &= \frac{1}{1 + (2\Omega/\kappa)^2}\,\hat{z}(t) - \frac{(2\Omega/\kappa)}{1 + (2\Omega/\kappa)^2}\,\hat{y}(t) \notag \\ 
   & = \sqrt{\eta_{\rm avg}}\,\left[\cos\theta\,\hat{z}(t) - \sin\theta\,\hat{y}(t)\right]. 
\end{align}
The averaging both attenuates the eigenvalue contrast of $\hat{z}$ by an efficiency factor $\eta_{\rm avg}$ and rotates the observable coupled to the resonator by an angle $\theta$ in the $yz$-plane. The tilt angle and efficiency are thus
\begin{align}
   &\theta = \arctan(\Omega \tau_c) = \arctan(2\Omega/\kappa)\\
   & \sqrt{\eta_{\rm avg}} = \cos\theta = \frac{1}{\sqrt{1 + (2\Omega/\kappa)^2}}.
   \label{eq:supp_tilt_and_meas_rate}
\end{align}
At longer times $t \gg \tau_c$ the geometric series in Eq.~\eqref{eq:qubita} then yields the standard steady state, but with a phase angle that depends upon the effective delay-averaged observable $\hat{z}_{\rm eff}(t)$ that is rotated by $\theta$ and with eigenvalues reduced by $\hat{z}_{\rm eff}^2 = \eta_{\rm avg}$. This tilt can be understood equivalently as two simultaneous measurements along $z$ and $y$ with differing measurement rates $\gamma_z = \eta_{\rm avg}\gamma_m \cos^2\theta$ and $\gamma_y = \eta_{\rm avg} \gamma_m \sin^2\theta$, that compete to rotate the effective measurement poles. 

In the main text we fit the tilt angle and measurement rate extracted from the LSTM to the expressions for $\theta$ and $\gamma_y + \gamma_z = \eta_\mathrm{avg} \gamma_m$ from Eq.~\eqref{eq:supp_tilt_and_meas_rate}. These results are shown in Fig.~\ref{fig:fig4}c.

\section{Derivation of measurement backaction maps with cavity effects}
\label{chap:supp_deriv_back_map}
To predict measurement backaction statistics shown in Fig.~\ref{fig:fig4}b, it is convenient to analyze an unnormalized Bloch-vector description $\vec{s} = (y,\,z,\,p)$ restricted to the $yz$-plane, where $p\in(0,1]$ is the total probability and the standard Bloch coordinates are obtained after renormalization, $\vec{S} = \mathcal{N}(\vec{s}) = (y/p,\,z/p)$. Focusing on the informational part of the measurement, a collected signal scaled to a measured observable $\hat{z}$ will have the form $r\,\mathrm{d}t = \langle \hat{z}\rangle\,\mathrm{d}t + \sqrt{\tau}\,\mathrm{d}W$ integrated over a small time bin $\mathrm{d}t$, where $\tau = 1/2\eta\Gamma_m$ is a characteristic timescale, $\Gamma_m$ is the induced ensemble-dephasing rate, $\eta\in[0,1]$ is the efficiency, and $\mathrm{d}W$ with $\langle \mathrm{d}W \rangle = 0$ and $\langle \mathrm{d}W^2\rangle = \mathrm{d}t$ is a Weiner-increment of zero-mean Gaussian noise. 

Observing such a signal $r$ induces a partial state collapse that has the form of a hyperbolic boost matrix \cite{Jordan2006,Patti2017},
\begin{align}
     \vec{s} &\xrightarrow{(r,\,\mathrm{d}t,\,\tau)} \hat{M}_z(r,\,\mathrm{d}t,\,\tau)\,\vec{s} \\
     \hat{M}_z(r,\,\mathrm{d}t,\,\tau) &\equiv \begin{bmatrix}1 & 0 & 0 \\ 0 & \cosh(r\,\mathrm{d}t/\tau) & \sinh(r\,\mathrm{d}t/\tau) \\ 0 & \sinh(r\,\mathrm{d}t/\tau) & \cosh(r\,\mathrm{d}t/\tau)\end{bmatrix},
\end{align}
Similarly, the measurement of a rotated observable  $\hat{z}(\theta)$ that is tilted by an angle $\theta$ in the $yz$-plane and further attenuated by an efficiency $\eta_2$ is readily obtained using a rotation matrix,
\begin{align}
     \hat{M}_{z(\theta)}(r,\,\mathrm{d}t,\,\tau,\,\eta_2) &\equiv \hat{R}_x(\theta)\hat{M}_z(r,\,\mathrm{d}t,\,\tau/\eta_2)\hat{R}^{-1}_x(\theta) \\
     \hat{R}_x(\theta) &\equiv \begin{bmatrix}\cos\theta & \sin\theta & 0 \\ -\sin\theta & \cos\theta & 0 \\ 0 & 0 & 1 \end{bmatrix}.
\end{align}
Computing the normalized state increment to linear order in $r\,dt$ yields the expected stochastic state increment,
\begin{align}\label{eq:stochinc}
    &\mathrm{d}\vec{S}(r,\,\mathrm{d}t,\,\tau,\,\eta_2,\,\theta) \equiv \mathcal{N}(\hat{M}_{z(\theta)}\,\vec{s} - \vec{s}) = \begin{bmatrix} \mathrm{d}y \\ \mathrm{d}z\end{bmatrix} \notag \\
    &=  \eta_2\,\frac{r\,\mathrm{d}t}{\tau}\begin{bmatrix} -yz\,\cos\theta + (1 - y^2)\sin\theta \\ -zy\,\sin\theta + (1-z^2)\cos\theta \end{bmatrix} + \mathcal{O}(r\,\mathrm{d}t)^2.
\end{align}
Ensemble-averaging the signal $r$ produces its mean $\langle r \rangle = \langle \hat{z}(\theta) \rangle = z\,\cos\theta + y\,\sin\theta$, yielding the mean backaction,
\begin{align}
    \langle \mathrm{d}\vec{S}\rangle &= \begin{bmatrix}\langle \mathrm{d}y \rangle \\ \langle \mathrm{d}z \rangle \end{bmatrix} \notag \\
    &\approx \eta_2\, \frac{\mathrm{d}t}{\tau}\,(z\,\cos\theta + y\,\sin\theta)\begin{bmatrix} -yz\,\cos\theta + (1 - y^2)\sin\theta \\ -zy\,\sin\theta + (1-z^2)\cos\theta \end{bmatrix},
\end{align}
from collapse, suppressing the distinct ensemble-dephasing terms appearing at second-order, $\langle(r\,\mathrm{d}t)^2\rangle = \tau\,\mathrm{d}t + \mathcal{O}(\mathrm{d}t^2)$.

To compute the variance of this informational backaction, it is convenient to decompose the random variable $r$ into a convex mixture, $r\,\mathrm{d}t = (+1\,\mathrm{d}t + \sqrt{\tau}\mathrm{d}W_+)p_+ + (-1\,\mathrm{d}t + \sqrt{\tau}\mathrm{d}W_-)(1-p_+)$ of two independent zero-mean Gaussian random variables, $\mathrm{d}W_+$ and $\mathrm{d}W_-$, conditioned on definite eigenvalues $\pm$ of $\hat{z}(\theta)$. Due to the linear prefactor of $r$ in Eq.~\eqref{eq:stochinc}, the variances can be readily computed using the conditional variance rule: the variance of the mixture is the mixture of the conditional variances plus the variance of the mixture means, $\text{Var}(Y) = \langle\text{Var}(Y|X)\rangle_X + \langle(Y|X)^2\rangle_X - (\langle Y|X\rangle_X)^2$. Computing $\text{Var}(r\, \mathrm{d}t)$, keeping the lowest order in $\mathrm{d}t$, and taking the square root then yields the standard deviation,
\begin{align}
    \text{Std}(\mathrm{d}\vec{S}) &\equiv \sqrt{\text{Var}(\mathrm{d}\vec{S})} \notag \\
    &\approx \sqrt{2\,\eta_2\,\frac{\mathrm{d}t}{\tau}}\,\begin{bmatrix} -yz\,\cos\theta + (1 - y^2)\sin\theta) \\ -zy\,\sin\theta + (1-z^2)\cos\theta \end{bmatrix}.
\end{align}
Note that the directionality of the standard deviation has been preserved when taking the square root.

Using the tilt angle $\theta = -\theta = -\arctan(2\Omega/\kappa)$ and the additional time-averaging efficiency factor $\eta_2 = \eta_{\rm avg} = \cos^2\theta = 1/(1 + (2\Omega/\kappa)^2)$ outlined in the previous section produces the variance vector plots in Fig.~\ref{fig:fig4}b of the main text.

%

\end{document}